\documentclass[%
amsmath,amssymb,
]{revtex4-2}

\usepackage{graphicx}
\usepackage{dcolumn}
\usepackage{bm}

\usepackage[utf8]{inputenc}
\usepackage[T1]{fontenc}
\usepackage{mathptmx}

\usepackage[a4paper]{geometry}
\usepackage{color}

\begin{document}
	
	
\title{Unusual chemical bond and spectrum of beryllium dimer in ground
$X^1\Sigma_g^+$ state}

\author{A. V. Mitin}
\address{Moscow Institute of Physics and Technology, 
Dolgoprudny, Moscow Region, 141700, Russia}
\address{Joint Institute for High Temperatures of RAS, 
125412 Moscow, Russia}
\email{mitin.av@mipt.ru.}

\author{A. A. Gusev}
\address{Joint Institute for Nuclear Research, Dubna, Moscow Region, 141980, 
	Russia}
\address{Dubna State University, Dubna, Moscow Region, 141980, Russia}
\email{gooseff@jinr.ru}	

\author{G. Chuluunbaatar}
\address{Joint Institute for Nuclear Research, Dubna, Moscow Region, 141980, 
	Russia}
\address{Dubna State University, Dubna, Moscow Region, 141980, Russia}
\email{chugal@jinr.ru}

\author{O. Chuluunbaatar}
\address{Joint Institute for Nuclear Research, Dubna, Moscow Region, 141980, 
	Russia}
\address{Institute of Mathematics and Digital Technology, Mongolian Academy of 
	Sciences, Ulaanbaatar, 13330, Mongolia}
\address{School of Applied Sciences, Mongolian University of Science and 
	Technology, Ulaanbaatar, 14191, Mongolia}
\email{chuka@jinr.ru}

\author{S. I. Vinitsky}
\address{Joint Institute for Nuclear Research, Dubna, Moscow Region, 141980, 
	Russia}
\address{Peoples' Friendship University of Russia (RUDN University),
	117198 Moscow, Russia}
\email{vinitsky@theor.jinr.ru.}

\author{V. L. Derbov} 
\address{N.G. Chernyshevsky Saratov National Research State University,
	Saratov, 410012, Russia}
\email{derbovvl@gmail.com}

\author{Le Hai Luong}
\address{Ho Chi Minh City University of Education, Ho Chi Minh City, Vietnam;}
\email{haill@hcmue.edu.vn; llhai611987@gmail.com}

\date{\today}
	
\begin{abstract}
This review outlines the main results which lead to understanding the dual
nature of the chemical bond in diatomic beryllium molecule in the ground
$X^1\Sigma_g^+$ state. It has been shown that the beryllium atoms are
covalently bound at low-lying vibrational energy levels ($\nu=0-4$), while at
higher ones ($\nu=5-11$) they are bound by van der Waals forces near the
right turning points. High precision \textit{ab initio} quantum mechanical
calculations of Be$_2$ resulted in the development of the modified expanded
Morse oscillator potential function which contains all twelve vibrational
energy levels [A.V. Mitin, Chem. Phys. Lett. 682, 30 (2017)]. The dual nature
of chemical bond in Be$_2$ is evidenced as a sharp corner on the attractive
branch of the ground state potential  curve. Moreover, it has been found that
the Douglas-Kroll-Hess relativistic corrections also show a sharp corner when
presented in dependence on the internuclear separation. The difference in
energy between the extrapolated and calculated multi-reference configuration
interaction energies in dependence on the internuclear separation also exhibits
singular point in the same region. The calculation of vibrational-rotational 
spectrum for the modified expanded Morse oscillator potential function and for 
function obtained with Slater-type orbitals [M. Lesiuk et al, Chem. Theory 
Comput. 15, 2470 (2019)] of the bound states of the beryllium dimer in the 
ground state was also considered in this review. Special attention was paid to 
the first calculations of the metastable vibrational-rotational complex-valued 
energy levels and the scattering length of the ground electronic state embedded 
in the continuum, along with the first theoretical estimations of the upper and 
lower border limits for the calculated vibrational-rotational energy levels of 
the bound as well as the metastable states. Such calculations are important for 
laser spectroscopy.
\end{abstract}
	
\maketitle


\section{Introduction}

The ground $X^1\Sigma_g^+$ state of the Be$_2$ molecule arises from the
interaction of two singlet states, which corresponds to the closed shells of
the Be atoms. Such a case occurs very rarely in diatomic molecules, which,
however, is attractive from the theoretical point of view. The \textit{ab
initio} quantum mechanical calculations of such molecular ground states become
significantly simple in comparison with other cases due to the simplification
of the configuration state functions (CSF). This fact, together with the small
number of electrons has led to numerous quantum mechanical calculations of the
ground state of the Be$_2$ molecule being performed beginning with the 30s of
the 20th century \cite{BartlettFurry_PhysRev_1931_38_1615}. Subsequent
investigations were reviewed in work \cite{RoeggenAlmlof_IJQC_1996_60_453}. The
potential energy curve of the ground state of the Be$_2$ molecule, calculated
and presented in publications \cite{RoeggenAlmlof_IJQC_1996_60_453,
ShirleyPetersson_CPL_1991_181_588, FustiMolnarSzalay_CPL_1996_258_400,
StarkMeyer_CPL_1996_258_421}, for the first time explicitly shows the
complicated character of the chemical bonding in Be$_2$. The potential curve
has a sharp corner at an internuclear distance of around 3.3 {\AA}. This has
been explained in 2009, both experimentally
\cite{MerittBondybeyHeaven_Science_2009_324_1548} and theoretically
\cite{Mitin_IntJQuantumChem_2011_111_2560} (the last article was submitted for
publication in November 2009 and published online in August 2010). Almost all
vibrational energy levels of the ground state were determined in the
experimental work and the Birge-Sponer \cite{BirgeSponer_PhysRev_1926_28_259}
dependence of $\Delta G_{\nu+1/2}= E_{\nu+1}-E_{\nu}$ as a function of the
vibrational quantum number $\nu$ was obtained. This dependence shows different
slopes of $\Delta G_{\nu+1/2}$ for the first five and for the upper vibrational
levels (see Figure 2 in work
\cite{MerittBondybeyHeaven_Science_2009_324_1548}). This means that the
vibrational constants of these groups of levels are different. On the other
hand, theoretical analysis of the leading configurations of the multi-reference
configuration interaction (MRCI) wave function
\cite{Mitin_IntJQuantumChem_2011_111_2560} shows that the chemical bonding
in Be$_2$ can be classified as a covalent bond at the internuclear distances
near the equilibrium point, while at larger distances it transforms to the van
der Waals interaction. Such a transformation of the electron density, which is
associated with a change of the type of chemical bonding in Be$_2$, has been
visualized in publication \cite{Mitin_CPL_2017_682_30}.

The Be$_2$ molecule is, thus, the first with a variable type of chemical bond
depending on the vibrational quantum number. In this connection, let us consider
a few representative recent \textit{ab initio} calculations of the dissociation
energy of Be$_2$ together with the corresponding values obtained from
experiments all of which are presented in Table \ref{Table_A1}.
\begin{table}[ht]
\caption{Dissociation energy $D_e$ (in cm$^{-1}$) of the ground state
$X^1\Sigma^+_g$ of the Be$_2$ molecule, calculated in a few selected recent
publications. Errors are shown, if estimated originally.}
\label{Table_A1}
	\begin{tabular*}{1.00\textwidth}{@{\extracolsep{\fill}}clcll}
	\hline
	Year &     Basis  &  Method      &     $D_e$    &      Reference      \\
	\hline
	2005 & 23s10p8d6f3g2h & Extended geminals& 945 $\pm$ 15 &
		R{\o}ggen, Veseth \cite{RoeggenVeseth_IJQC_2005_101_201} \\
	2007 & aug-cc-pcvqz+bf  & CCSD(T)+FCI/CBS  & 938.7 $\pm$ 15 & Patkovski
		\textit{et al.} \cite{PPS_JPCA_2007_111_12822} \\
	2009 & \textit{ab initio} & morphed RPC & 934.6 & Patkowski \textit{et al.}
		\cite{PatkowskiSpirkoSzalewicz_Science_2009_324_1548} \\
	2010 & aug-cc-pvqz  & FCI/CBS/correl corr  & 911.7 & Schmidt
		\textit{et al.} \cite{SIR_JPCA_2010_114_8687} \\
	2011 & aug-cc-pv6z  & CCSD(T)/FCI+corr  & 935.1 $\pm$ 10 & Koput
		\cite{Koput_PCCP_2011_13_20311} \\
	2014 &     &    DMRG      & 931.2        & Sharma \textit{et al.}
		\cite{SYBUC_JChemPhys_2014_140_104112} \\		
	2015 & STO/atc-etcc-6 & fc/ae FCI/CCSD(T)/CBS & 929.0 $\pm$
		1.9 & Lesiuk \textit{et al.} \cite{LPMJM_PhysRevA_2015_91_012510} \\
	2017 & t-aug-ccpV6Z & MRCI/CBS & 929.8 & Mitin
		\cite{Mitin_CPL_2017_682_30} \\
	2018 & aug-cc-pVQZ & CCSDTQ+corr& 928.0 $\pm$ 1.9 & Magoulas
		\textit{et	al.} \cite{MBSP_JPhysChem_2018_122_1350} \\
	2018 & cc-pV6Z & FCI/CBS+corr& 922.9 $\pm$ 1.9 & Rolik
		\textit{et	al.} \cite{RolikKallay_JChemPhys_2018_148_124108} \\
	2019 & STO/atc-etcc-6 & FCI/CBS+corr& 934.6 $\pm$ 2.5 & Lesiuk
		\textit{et	al.} \cite{LPBMM_JChemTheorComp_2019_15_2470} \\
	2009 & Experiment & EMO & 929.7 $\pm$ 2.0& Merrit \textit{et al.}
		\cite{MerittBondybeyHeaven_Science_2009_324_1548} \\
	2014 & Experiment & EMO & 929.6 & Meshkov \textit{et al.}
		\cite{MSHHLR_JCP_2014_140_064315} \\				
	2014 & Experiment & MLR & 934.8 & Meshkov \textit{et al.}
		\cite{MSHHLR_JCP_2014_140_064315} \\	
	2017 & Experiment & MEMO & 929.7 $\pm$ 2.0 & Mitin
	\cite{Mitin_CPL_2017_682_30} \\
	\hline
	\end{tabular*}
\end{table}
An analysis of these results shows that the values of $D_e$ are close to each
other when they were computed directly as a difference of the total energies at
the equilibrium point and at the dissociation limit or determined as parameter
$D_e$ in the expanded Morse oscillator (EMO) potential function, defined as
\cite{CoxonHajigeorgiou_JMS_1990_139_84}:
\begin{equation}
	V(r)=D_e[1-e^{-\phi(r)(r-r_e)}]^2    \label{eq_01} \; ,
\end{equation}
\begin{eqnarray}
	\phi(r)=\sum_{i=0}^n\phi_i\left(\frac{r^p-r_{ref}^p}{r^p+r_{ref}^p}\right)^i
	\label{eq_02},
\end{eqnarray}
which is used for approximation of the vibrational energy levels or points of
the potential curve. Two alternate values of $D_e$ predominate in these
results: values of $D_e$ of about 930 cm$^{-1}$ were obtained in publications
\cite{SYBUC_JChemPhys_2014_140_104112, LPMJM_PhysRevA_2015_91_012510,
Mitin_CPL_2017_682_30, MBSP_JPhysChem_2018_122_1350,
MerittBondybeyHeaven_Science_2009_324_1548, MSHHLR_JCP_2014_140_064315}. On the
other hand, values of about 935 cm$^{-1}$ have been obtained when the
approximated potential functions, for example EMO, were corrected by the
addition of the long-range asymptotic correction in the form of inverse-power
terms and were used for fitting the experimental energy levels or the
calculated theoretical points of a potential curve. This was the case in works
\cite{PatkowskiSpirkoSzalewicz_Science_2009_324_1548,
MSHHLR_JCP_2014_140_064315, LPBMM_JChemTheorComp_2019_15_2470}.
It is especially important to note that the values of $D_e$ equaling 929.6 and
934.8 cm$^{-1}$ were obtained in work \cite{MSHHLR_JCP_2014_140_064315} when
the same experimentally determined vibration energy levels were approximated by
using EMO and Morse long-range (MLR) potential functions, i.e. without a
correction for long range asymptotic behavior and with it. Additionally, in the
same article, the experimental energy levels were fitted using the Chebyshev
polynomial expansion (CPE) function. In general, the use of orthogonal
polynomials, including Legendre and Chebyhev polynomials of the first and the
second kind, for the approximation of a potential curve by the optimal
approximating polynomial was proposed in works
\cite{Mitin_ZhFizKhim_1990_64_2041, Mitin_JCC_1998_19_94}.
These three types of polynomials permit the assignment of different weights for
approximated points, which might be useful for accounting for variable
experimental measurement errors more correctly.

The inconsistency of the $D_e$ values determined by the different ways
mentioned above can be understand if note that the potential curves are
usually defined by minimizing the root mean square (RMS) deviation of the
optimized potential curve from the \textit{ab initio} calculated points on it or
similarly by minimizing the RMS deviation of the calculated vibrational energy
levels from the experimental ones. This means that for selected formula 
describing a potential curve only the full set of parameters in formula have a 
sense. The different formulas usually have different number of parameters. 
Therefore, individual comparison the values of parameters from the different 
sets of parameters, in particular $D_e$, has no meaning. 

Obviously, this case shows a problem of inconsistency between physics and 
mathematics, because parameter $D_e$ is usually considered as a dissociation 
energy in physics. But they are differ in different formulas. For this reason, 
an explicit reference on formula for $D_e$ would be useful in discussions for 
clarity. Although, it is well known that the dissociation energies can not be 
directly experimentally measured and $D_0$ can only be approximately estimated 
from the vibrational-rotational spectra \cite{Gaydon_ProcPhysSoc_1946_58_525}.

The dual nature of the chemical bonding in Be$_2$ is evident in the
calculations of the vibrational energy levels of the ground state with the EMO
potential function. The transition to the twelfth vibration level, which was
clearly observed in the spectrum, has not been identified in the experimental
work \cite{MerittBondybeyHeaven_Science_2009_324_1548}. This happens because
the EMO potential function used to calculate the vibrational energy levels in
Be$_2$ holds only eleven energy levels. However, the EMO potential function was
originally proposed for the description of the covalent chemical bond. On the
other hand the van der Waals bond is weaker and more important for larger
internuclear distances in comparison to the covalent bond. For this reason, it
is clear that the EMO potential function of Be$_2$ is more narrow near
dissociation limit in comparison with the correct one, which has to describe
the van der Waals interaction at large distances in Be$_2$. This was noted
first in work \cite{PatkowskiSpirkoSzalewicz_Science_2009_324_1548}. Later,
this problem of the EMO potential function
\cite{MerittBondybeyHeaven_Science_2009_324_1548}
was corrected in the modified EMO (MEMO) potential function by including the
\textit{ab initio} MRCI potential near the dissociation limit in the EMO
potential function \cite{Mitin_CPL_2017_682_30}. The MEMO potential function,
constructed in such a way, holds all twelve vibrational energy levels of
Be$_2$.

An important question follows from the dual nature of the chemical bond in the
Be$_2$ molecule. Most of the available Gaussian basis sets for \textit{ab
initio} calculations were developed for the description of the covalent and
ionic type of bonding. However, due to the dual nature of the chemical bond in
Be$_2$, the basis set used in \textit{ab initio} calculations have to describe
both covalent bonding in the vicinity of equilibrium distance and van der Waals
bonding at large distances at the same level of quality.

The significance of this problem can be demonstrated by considering the
convergence of the calculated results depending on the size of the employed
basis sets, i.e. the number of basis functions. In this connection, a few large
Gaussian basis sets have been developed and used in extended quasi-relativistic
MRCI calculations of the dissociation energy $D_e$ of the Be$_2$ molecule. This
energy was estimated as a difference of the total energies computed at point
near the equilibrium and at point corresponding to the dissociation limit, i.e.
at $R=2.44$ and $R=80.0$ \AA, correspondingly. For the first time, the
dependence of $D_e$ on the basis set size is given in Table \ref{Table_A2}. The
presented values of $D_e$ were corrected for the basis set superposition error
(BSSE).
\begin{table}[ht]
	\caption{Dissociation energies $D_e$ (in cm$^{-1}$) of the ground state
	$X^1\Sigma^+_g$ of Be$_2$ calculated using the MRCI method together with
	the corresponding basis sets and the number of CSF used in calculation.}
	\label{Table_A2}
	\begin{tabular*}{1.00\textwidth}{@{\extracolsep{\fill}}ccc}
	\hline
	Basis &  CSF  &    $D_e$         \\
	\hline
	(16s10p5d4f3g2h1i)/[7s6p5d4f3g2h1i]+(3s3p3d3f3g3h3i) & 200 073 854 & 948.8\\
	(18s12p5d4f3g2h1i)/[8s7p5d4f3g2h1i]+(3s3p3d3f3g3h3i) & 203 636 850 & 937.7\\
	(18s12p6d5f4g3h2i)/[8s7p6d5f4g3h2i]+(3s3p3d3f3g3h3i) & 217 677 644 & 934.0\\
	(20s14p6d5f4g3h2i)/[9s8p6d5f4g3h2i]+(3s3p3d3f3g3h3i) & 242 592 786 & 930.4\\
	(22s16p6d5f4g3h2i)/[10s9p6d5f4g3h2i]+(3s3p3d3f3g3h3i)& 246 197 719 & 928.0\\
\hline
\end{tabular*}
\end{table}

It is well known that the contribution of the van der Waals interaction to
$D_e$ is smaller in comparison with the covalent one. In this connection it is
reasonable to assume that the correlation energy corresponding to the van der
Waals interaction converge to their limit faster with increasing of the number
of basis functions and CSF in comparison with that for the covalent
interaction. The results presented in this Table \ref{Table_A2} shows that the
$D_e$ reduces from over 935 to below 929.6 cm$^{-1}$, which is the lowest
experimental value of Be$_2$ dissociation energy. Hence, one can conclude that
the fraction of correlation energy which was taken into account in the
calculations presented in this table reduces with increasing the number of
basis functions despite of small enlarging the number of CSF. Therefore, an
equivalent description of the van der Waals and covalent bonding in Be$_2$ was
not reached in these calculations despite large number of CSFs included in the
MRCI calculations \cite{MitinMMG2014_1_2_10}.

It should be noted that quantum mechanical \textit{ab initio} calculations are
based on the statement that the obtained results must converge to the correct
ones when increasing the number of employed basis functions and CSFs. It is
well known that all published \textit{ab initio} calculations of Be$_2$ were
only performed with a single selected basis set (or with two sets when
extrapolation to the infinite number of basis functions was employed). This
means that a good agreement in some publications between the calculated $D_e$
and the experimental ones must be considered just as a lucky case.

\section{\textit{Ab initio} calculations of Be$_2$ and modified EMO potential
energy function} \label{section2}

In the light of the comments presented above, the MRCI calculation of the
potential curve of the ground $X^1\Sigma^+_g$ state of the Be$_2$ molecule
given in publication \cite{Mitin_CPL_2017_682_30} looks like a compromise in
accounting for the outlined problems which, however, reproduce the experimental
dissociation energy $D_e$ well.

High-precision \textit{ab initio} MRCI calculations of the Be$_2$ potential
curve have been performed using a well known program \cite{MOLPRO_2003} with
correlation consistent cc-pVQZ and cc-PV5Z basis sets
\cite{Dunning_JCP_1989_90_1007}, extended by four augmented functions
\cite{KDH_JCP_1992_96_6796, WoonDunning_JCP_1994_100_2975} for the better
description of the van der Waals interaction. Molecular orbitals obtained in
the Hartree-Fock calculations were transformed into pseudo-natural orbitals to
improve convergence of the configuration expansion set. Relativistic effects
have been taken into account using the Douglas-Kroll-Hess approach
\cite{DouglasKroll_AnnPhys_1974_82_89, Hess_PRA_1985_32_756,
Hess_PRA_1986_33_3742}. The calculated total electronic energies have been
corrected using the Boys-Bernardi counterpoise method
\cite{BoysBernardi_MolPhys_1970_19_553} to eliminate the BSSE
\cite{vDvDvdRvL_ChemRev_1994_94_1873} and then were extrapolated to the
infinite basis set for the case of using natural orbitals in accordance with
the considerations given in work \cite{Gdanitz_JCP_2000_113_5145}. The scalar
relativistic corrections and application of extrapolation to the infinite basis
set in the Be$_2$ molecule were investigated in Refs.
\cite{Mitin_IntJQuantumChem_2011_111_2560, Koput_PCCP_2011_13_20311,
PeterssonShirley_CPL_1989_160_494, ShirleyPetersson_CPL_1991_181_588,
LPBMM_JChemTheorComp_2019_15_2470}. The total many configuration molecular wave
function for the largest MRCI calculation of Be$_2$ was constructed from the
list of reference configurations consists of a complete active space of about
$1.6\cdot10^4$ configurations formed by 24 molecular orbitals of $D_{2h}$
symmetry, which describe the excitations of electrons to the $2p$, $3s$, $3p$,
and higher orbitals of Be. All singly and doubly excited configurations
generated from this list of reference configurations have been included in the
MRCI calculations. The largest calculation includes about $37.2\cdot10^6$
configuration state functions.

The values of the total potential energy of Be$_2$ calculated at 77 points of
internuclear separation obtained in these calculations are presented in Table
\ref{Table_A3}.
\begin{table}[ht]
\caption{MRCI potential energy curve of the $X^1\Sigma ^+_g$  state of Be$_2$.
Internuclear separation is in angstrom (\r{A}), the total energy is in
(cm$^{-1}$).}
\label{Table_A3}
	\begin{tabular*}{1.00\linewidth}{@{\extracolsep{\fill}}rrrrrrrr}
	\hline
	R & E$_{tot}$ & R & E$_{tot}$ & R & E$_{tot}$ & R & E$_{tot}$ \\
	\hline
	1.50& 24554.9792& 3.00& -478.4235&  6.00& -29.8545& 24.00& -0.0275 \\
	1.60& 16238.9043& 3.10& -415.5393&  6.50& -17.9725& 25.00& -0.0250 \\
	1.70& 10237.2474& 3.20& -366.2914&  7.00& -11.1156& 26.00& -0.0250 \\
	1.80&  6045.0232& 3.30& -328.1793&  7.50&  -7.0865& 27.00& -0.0217 \\
	1.90&  3201.9108& 3.40& -298.4670&  8.00&  -4.6482& 28.00& -0.0136 \\
	2.00&  1344.4657& 3.50& -274.7081&  9.00&  -2.1839& 29.00& -0.0141 \\
	2.10&   192.0413& 3.60& -254.8773& 10.00&  -1.1249& 30.00& -0.0089 \\
	2.20&  -469.1904& 3.70& -237.3498& 11.00&  -0.6198& 32.00& -0.0160 \\
	2.30&  -800.1700& 3.80& -220.9879& 12.00&  -0.3804& 34.00& -0.0099 \\
	2.40&  -920.4464& 3.90& -205.7910& 13.00&  -0.2399& 36.00& -0.0086 \\
	2.42&  -927.3053& 4.00& -192.2289& 14.00&  -0.1588& 38.00& -0.0062 \\
	2.43&  -929.0688& 4.20& -167.4891& 15.00&  -0.1080& 40.00& -0.0072 \\
	2.44&  -929.8058& 4.40& -143.7740& 16.00&  -0.0818& 42.00& -0.0052 \\
	2.45&  -929.5907& 4.60& -121.5195& 17.00&  -0.0658& 44.00& -0.0030 \\
	2.46&  -928.4713& 4.80& -101.3230& 18.00&  -0.0448& 46.00& -0.0028 \\
	2.50&  -916.0723& 5.00&  -83.5874& 19.00&  -0.0395& 48.00& -0.0026 \\
	2.60&  -846.1587& 5.20&  -68.4332& 20.00&  -0.0318& 50.00&  0.0000 \\
	2.70&  -749.2164& 5.40&  -55.7499& 21.00&  -0.0227&      & \\
	2.80&  -648.4697& 5.60&  -45.2901& 22.00&  -0.0262&      & \\
	2.90&  -556.4006& 5.80&  -36.7651& 23.00&  -0.0343&      & \\
	\hline
	\end{tabular*}
\end{table}
A~comparison of the theoretical potential energy function given in this table
with the EMO potential function derived from fitting the experimental energy
levels shows that, near the equilibrium point, the EMO function is broader
compared with the theoretical potential energy function. However, the EMO
function is narrower in comparison to the theoretical one near the dissociation
limit, which is expected in light of the above analysis. For this reason, the
EMO potential function has only eleven  vibration levels, while the twelfth
energy level is pushed out to the continuum spectrum.

From this analysis follows that the EMO potential function derived from the
experimental results can be improved by modifying its part near dissociation
limit. Following this conclusion, the modified MEMO potential function was
constructed by replacing the repulsive branch of the EMO potential function
above the dissociation limit and its attractive branch above 895 cm$^{-1}$ with
the theoretical counterparts.

The parameters of the MEMO potential function obtained in such a way and the
corresponding vibrational energy levels are given in Table \ref{Table_A4}. The
values of the vibrational energy levels calculated with other published
potentials, where all twelve vibration energy levels were obtained, are also
given in this table. A comparison of the EMO and MEMO potential functions shows
that the modified function has all twelve vibrational levels and better
describes the experimental vibrational energy levels due to smaller RMS error.
The twelfth level of MEMO potential has three rotational levels with $J=0,1,2$
which are about 0.3 cm$^{-1}$ below the dissociation limit.

\begin{turnpage}
\begin{table*}
\caption{Spectroscopic parameters of the Be$_2$ molecule. Vibration energy
levels are in cm$^{-1}$, $R_e$ is given in \AA.}
\label{Table_A4}
\begin{tabular*}{\linewidth}{@{\extracolsep{\fill}}ccccccccccccccc}
\hline
 & MRCI& MEMO& MEMO* & MRCI& MRACPF& SAPT& CV+F+R& CCSDTQ& FCI/CBS&
 STO* & morphed& MRL& EMO & Exp\\
 &     &     &     &       &     &       &   +corr& +corr& RPC&  &  &  & & \\
Ref. & \cite{Mitin_CPL_2017_682_30} & \cite{Mitin_CPL_2017_682_30} &
\cite{DerbovJQSRT2021_262_107529} &
\cite{StarkMeyer_CPL_1996_258_421} & \cite{Gdanitz_CPL_1999_312_578} &
\cite{PatkowskiSpirkoSzalewicz_Science_2009_324_1548}  &
\cite{Koput_PCCP_2011_13_20311} & \cite{MBSP_JPhysChem_2018_122_1350} &
\cite{LPBMM_JChemTheorComp_2019_15_2470} &
\cite{DerbovJQSRT2021_262_107529} &
\cite{PatkowskiSpirkoSzalewicz_Science_2009_324_1548} &
\cite{MSHHLR_JCP_2014_140_064315} &
\cite{MerittBondybeyHeaven_Science_2009_324_1548} &
\cite{MerittBondybeyHeaven_Science_2009_324_1548}  \\
\hline
$R_e$     &2.4427&2.4535&2.4534&2.4485& 2.4443  & 2.443& 2.434&2.4344& 2.4436&
2.447 &2.445 & 2.438& 2.4536&    \\
$D_e$     &929.84&929.74& 929.8     &   893&898$\pm$8&938.7 &935.1$\pm$10&928.0
& 934.6$\pm$2.5& 934.4     &934.6 & 934.8    &929.74& 929.7$\pm$2.0 \\
$D_0$     &802.59& 806.5&806.0      & 768.2& 772.2   &      &808.3       &798.5
&807.7         &807.7      &      & 808.2    & 806.5& 807.4         \\
G(0)      & 127.2& 123.3& 123.7& 124.8& 125.8&      & 126.8& 129.5&      &
126.6&      & 126.7&      &       \\
G(1)-G(0) & 224.0& 222.2& 222.5& 218.4& 218.6& 222.3& 222.7& 227.4& 223.4&
223.5& 222.6& 222.9& 222.7& 222.6 \\
G(2)-G(0) & 400.3& 397.6& 397.3& 387.0& 388.1& 397.6& 396.8& 405.1& 400.1&
398.2& 397.0& 397.4& 397.8& 397.1 \\
G(3)-G(0) & 523.0& 517.9& 517.7& 499.0& 503.2& 520.3& 517.8& 526.0& 517.3&
519.3& 518.2& 518.4& 518.2& 518.1 \\
G(4)-G(0) & 600.1& 595.1& 594.8& 568.0& 576.0& 597.9& 594.7& 598.5& 595.1&
595.7& 594.8& 595.1& 595.4& 594.8 \\
G(5)-G(0) & 655.9& 652.1& 651.9& 622.4& 631.0& 655.1& 651.6& 652.6& 651.7&
652.2& 651.5& 651.8& 652.4& 651.5 \\
G(6)-G(0) & 701.1& 699.1& 698.9& 668.5& 676.2& 702.6& 698.9& 697.9& 698.7&
699.3& 698.7& 699.0& 699.4& 698.8 \\
G(7)-G(0) & 738.7& 738.0& 737.7& 706.1& 712.4& 741.7& 738.0& 735.3& 738.0&
738.1& 737.6& 738.0& 738.2& 737.7 \\
G(8)-G(0) & 767.9& 768.6& 768.2& 734.8& 740.0& 772.4& 768.6& 764.1& 768.3&
768.6& 768.2& 768.5& 768.8& 768.2 \\
G(9)-G(0) & 788.0& 790.0& 789.7& 754.2& 758.9& 794.3& 790.4& 783.9& 790.1&
790.1& 790.0& 790.2& 790.7& 789.9 \\
G(10)-G(0)& 799.0& 802.1& 801.6& 764.6& 769.0& 807.1& 803.1& 794.8& 802.6&
802.6& 802.5& 802.8& 803.4& 802.6 \\
G(11)-G(0)& 802.4& 806.2& 805.7& 768.0& 772.0& 811.9& 807.9& 798.2& 807.5&
807.2& 807.1& 807.6&      &       \\
RMS       &   3.1&$<$0.4&$<$0.4&  25.7&  22.6&  3.4 & 0.3  &      &  1.0 &
  0.7&$<$ 0.1&  0.3&   0.6&       \\
\hline
\end{tabular*}
\end{table*}
\end{turnpage}

The \textit{ab initio} potentials presented in publications
\cite{PatkowskiSpirkoSzalewicz_Science_2009_324_1548, Koput_PCCP_2011_13_20311}
have RMS errors of about 0.1 and 0.3 cm$^{-1}$, correspondingly, while the
potential obtained in work \cite{LPBMM_JChemTheorComp_2019_15_2470} has a
somewhat larger RMS error of about 1.0 cm$^{-1}$. The values of $D_e$ for these
three potentials, equal to 934.6, 935.1 and 934.6 cm$^{-1}$, are close to each
other. Although, only for the second potential, $D_e$ probably was estimated as
the difference of the total energy at the equilibrium point and at the distance
near dissociation limit, while for the first and third potentials $D_e$ were
determined by fitting the theoretical points calculated in limited regions by
potentials, which takes into account long-range asymptotic corrections.

Other differences between potentials can be observed by comparing the
deviations of calculated vibrational energies from the experimental ones
presented in Table \ref{Table_A5}. For the morphed RPC  
\cite{PatkowskiSpirkoSzalewicz_Science_2009_324_1548}, MRL  
\cite{MSHHLR_JCP_2014_140_064315} and MEMO \cite{Mitin_CPL_2017_682_30} 
potentials deviations are distributed approximately uniformly over the whole 
range of energy levels. For the CV+F+R potential 
\cite{Koput_PCCP_2011_13_20311}, the largest deviations are located close to 
the bottom and top of the vibrational energy spectrum, while for the 
FCI/CBS+corr potential \cite{LPBMM_JChemTheorComp_2019_15_2470} the largest 
deviations are observed only for low lying vibration energy levels. Such 
variations in the distributions of deviation errors can be explained as being 
the manifestation of the problem mentioned above: the equivalent description of 
the covalent and van der Waals chemical bonding in Be$_2$ by the basis set used 
in \textit{ab initio} calculations. 

\begin{table*}
	\caption{Deviations of the calculated vibrational energy levels from the
	experimental ones ($E(exp)-E(calc)$ for different vibration number $\nu$ in
	cm$^{-1}$.}
	\label{Table_A5}
	\begin{tabular*}{\linewidth}{@{\extracolsep{\fill}}cccccccc}
		\hline
		     & morphed & CV+F+R & FCI/CBS & MRL & STO* & MEMO & MEMO* \\
             &         &	    & +corr   &     &      &      &       \\
 	    $\nu$~/~Ref. &\cite{PatkowskiSpirkoSzalewicz_Science_2009_324_1548}  &
        \cite{Koput_PCCP_2011_13_20311} &
        \cite{LPBMM_JChemTheorComp_2019_15_2470} &
        \cite{MSHHLR_JCP_2014_140_064315} &
        \cite{DerbovJQSRT2021_262_107529} &
        \cite{Mitin_CPL_2017_682_30} &
        \cite{DerbovJQSRT2021_262_107529} \\
		\hline
		 1 &  0.0 & -0.1 & -0.8 & -0.3 & -0.9 &  0.4 &  0.1 \\
		 2 &  0.1 &  0.3 & -3.0 & -0.3 & -1.1 & -0.5 & -0.2 \\
		 3 & -0.1 &  0.3 &  0.8 & -0.3 & -1.2 &  0.2 &  0.4 \\
		 4 &  0.0 &  0.1 & -0.3 & -0.3 & -0.9 & -0.3 &  0.0 \\
		 5 &  0.0 & -0.1 & -0.2 & -0.3 & -0.7 & -0.6 & -0.4 \\
		 6 &  0.1 & -0.1 &  0.1 & -0.2 & -0.5 & -0.3 & -0.1 \\
		 7 &  0.1 & -0.3 & -0.3 & -0.3 & -0.4 & -0.3 &  0.0 \\
		 8 &  0.0 & -0.4 & -0.1 & -0.3 & -0.4 & -0.4 &  0.0 \\
		 9 & -0.1 & -0.5 & -0.2 & -0.3 & -0.2 & -0.1 &  0.2 \\
		10 &  0.1 & -0.5 &  0.0 & -0.2 & 0.0 &  0.5 &  1.0 \\
		\hline
	\end{tabular*}
\end{table*}

A comparison of the vibrational energy levels calculated for EMO/MEMO and
MLR/CPE potentials shows that the main discrepancy between them arises from the
different values of the zero vibrational energy level $G(0)$: 123.2 and 123.3
cm$^{-1}$ for EMO/MEMO, while for MLR/CPE potentials the values of G(0) are
126.6 and 126.8 cm$^{-1}$, correspondingly. This difference gives the main
contribution to the different values of $D_e$ equal to 929.7 and 934.8
cm$^{-1}$, respectively. In general, the observed significant influence of the
values of a potential at large distances on $G(0)$ is an unexpected result,
which can not be easily explained. This is a problem for further investigations.

In this connection, note that the good agreement of the experimental value of
$G(0)$ for MLR/CPE potentials with the theoretical one obtained in work
\cite{Koput_PCCP_2011_13_20311} - 126.9 cm$^{-1}$ can not be considered
indicative of a correct result. This result was obtained using a basis set with
contracted $s$ and $p$ functions, however when a basis set with with
uncontracted $s$ and $p$ functions is used, as was done in work
\cite{StarkMeyer_CPL_1996_258_421}, a value of $G(0)$ of 124.8 cm$^{-1}$ is
obtained. This points out that the contraction of $s$ and $p$ functions in
basis sets probably noticeably reduces the basis set flexibility in describing
the potential near the minimum point.

Summarizing the above notes we can conclude that the porphed, MRL and MEMO 
potentials are the best ones among those derived from the experimental data, 
although the differences in value of $G(0)$ and $D_e$ are now not explainable. 
Theoretical potentials presented in publications
\cite{PatkowskiSpirkoSzalewicz_Science_2009_324_1548, Koput_PCCP_2011_13_20311}
stand out favorably from the known ones. Most importantly, however, special
attention should be paid to comparing different forms of employed asymptotic
corrected potentials.

\section{Chemical bonding in beryllium dimer and experimental potential energy
function}

\begin{figure}[h]
	\begin{center}
		\includegraphics[width=0.9\textwidth]{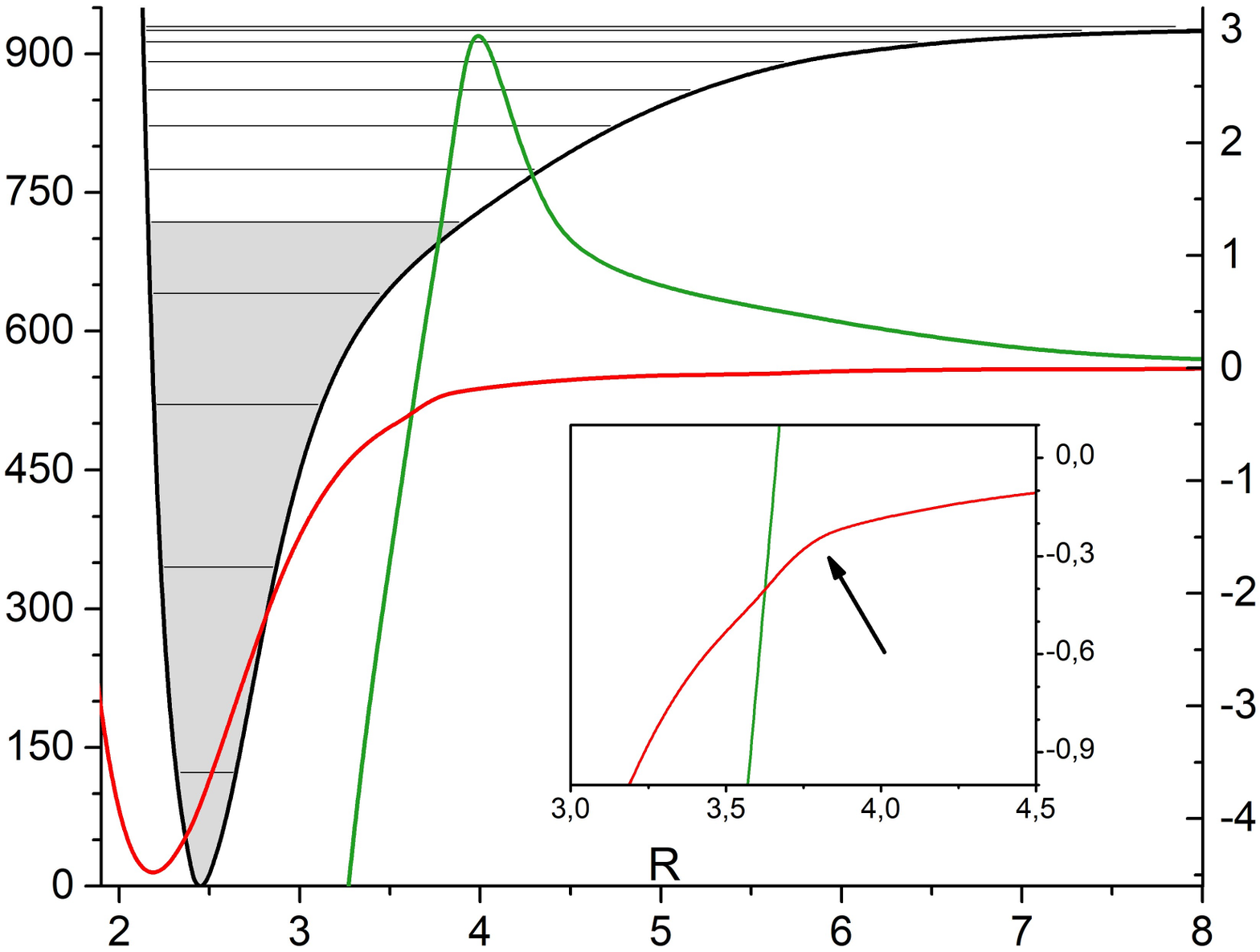}
	\end{center}
	\caption{The $X^1\Sigma^+_g$ ground state Be$_2$ MEMO potential energy curve
	together with the calculated vibrational energy levels (black)
	corresponding to the left vertical axis. The relativistic corrections
	(red) and the differences between extrapolated and calculated MRCI
	energies (green) corresponding to the right vertical axis as a function of
	the internuclear separation (R) given in \AA. Energies are given in
	cm$^{-1}$.}
	\label{Figure_1}
\end{figure}

The $X^1\Sigma^+_g$ ground state MEMO potential curve of Be$_2$, constructed in
work \cite{Mitin_CPL_2017_682_30}, is presented on Figure \ref{Figure_1}
together with the calculated vibrational energy levels, designated by
horizontal lines at corresponding energies. The figure shows that the slope of
the attractive part of the potential curve changes after the fourth vibrational
level, indicating the change of the character of the chemical bond in Be$_2$:
having a predominantly covalent nature in the low part (shaded in gray), and a
more strongly van der Waals bond for the higher vibrational levels.  This means
that the low-lying ($\nu=0-4$) and upper vibrational energy levels can only be
correctly described by two separate sets of Dunham coefficients
\cite{Dunham_PhysRev_1932_41_713,Dunham_PhysRev_1932_41_721}, one of which is
used for describing the low-lying levels and the other one for the upper ones,
i.e. the fundamental vibration frequencies $\omega_e$ (Y$_{10}$), or force
constants of these vibration levels, are different. As mentioned above, this
fact has been experimentally shown in work
\cite{MerittBondybeyHeaven_Science_2009_324_1548}, where the dependence of
$\Delta G_{\nu+1/2}=E_{\nu+1}-E_{\nu}$ as a function of the vibrational
quantum number $\nu$ was constructed. Figure 2 in this reference displays
different slopes of $\Delta G_{\nu+1/2}$ for the first five vibrational levels
compared to the higher lying ones. This leads to the conclusion that the
chemical bonding in Be$_2$ on the low-lying and upper vibrational energy levels
are different. This conclusion has been confirmed by an investigation of the
expansion coefficients of the MRCI wave functions, published in work
\cite{Mitin_IntJQuantumChem_2011_111_2560}, which shows that the distribution
of the total electron density in Be$_2$ corresponds to the usual covalent
interaction at low-lying vibrational levels near the equilibrium point. On the
other hand, for the upper levels, near the right turning points, the total
electron density distributions are described as two asymmetric ellipsoidal
distributions (pointed at each other) of the electron densities centered on the
positions of the nuclei. This type of total electron density distribution
clearly corresponds to the van der Waals interaction.

\begin{figure}
	\begin{center}
		\includegraphics[scale=0.35]{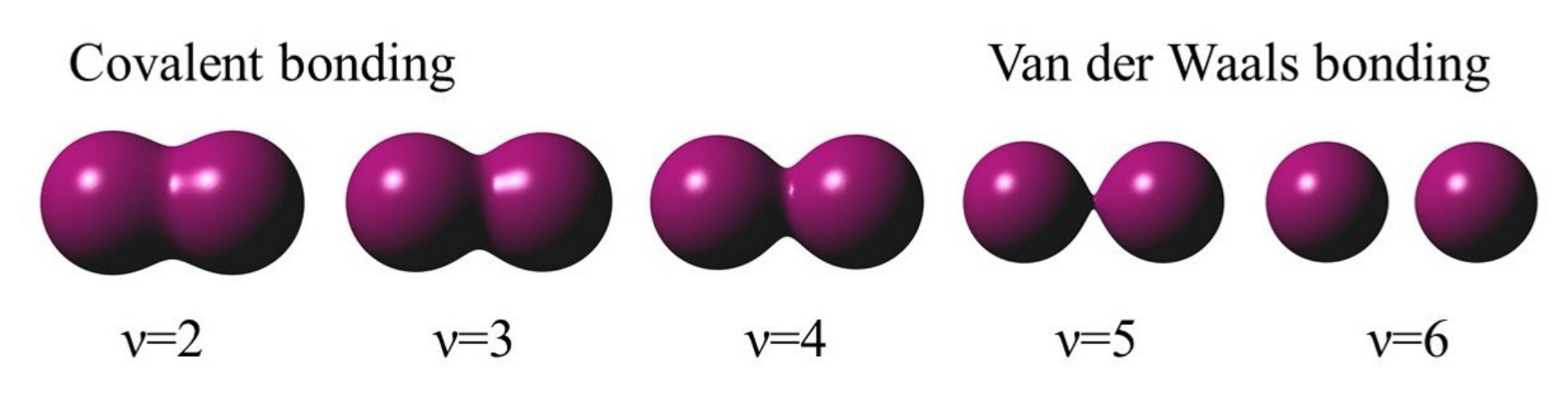}
	\end{center}
	\caption{The transformation of the total electron density isosurfaces for
	the Be$_2$ molecule at the right turning points for different vibrational
	quantum numbers $\nu$, showing the modification of the covalent bonding at
	low-lying $\nu$ into the van der Waals one with increasing~$\nu$.}
	\label{Figure_2}
\end{figure}

In this connection, to show dual nature of chemical bonding in the beryllium
dimer, the total electron densities of Be$_2$ were calculated in work
\cite{Mitin_CPL_2017_682_30} for internuclear separations corresponding to the
right turning points of the vibrational levels $\nu=2-6$ and are presented on
Figure \ref{Figure_2}. This figure visualizes the successive transformation of
the covalent bonding at the right turning points, which is realized in Be$_2$
on the low-lying vibrational levels $\nu=0-4$, to the van der Waals bonding on
the upper vibrational levels $\nu=5-11$.

The transformation of the chemical bonding in Be$_2$ depending on the
vibrational quantum number or the main configurations of the total
molecular wave function depending on the internuclear separation should be
evident in other physical characteristics of the molecule as well. In
particular, the investigations of the relativistic corrections
\cite{DouglasKroll_AnnPhys_1974_82_89, Hess_PRA_1985_32_756,
Hess_PRA_1986_33_3742} as a function of the internuclear separation, presented
in work \cite{Mitin_Thesis_2013}, shows that this dependence also has a sharp
turn indicated by arrow in the inset on Figure \ref{Figure_1}. The dependence
of the difference of energies between the extrapolated MRCI and the MRCI energy
obtained with the cc-pV5Z basis sets in dependence on the internuclear distance
\cite{Mitin_Thesis_2013} presented on Figure \ref{Figure_1} also has a specific
point in the same region of the internuclear separations. This is especially
interesting because only the energies obtained with two different basis sets
are included in the formula of the employed extrapolation method
\cite{Gdanitz_JCP_2000_113_5145}.

In the experimental work \cite{MerittBondybeyHeaven_Science_2009_324_1548}, the
dual nature of the chemical bond in Be$_2$ was overlooked. For this reason
the usual EMO potential function has been employed by the authors for the
description of the interaction potential in it. However, the Morse oscillator
(MO) potential function introduced in work \cite{Morse_PR_1929_1_57} and its
generalization - expanded Morse oscillator EMO function
\cite{CoxonHajigeorgiou_JMS_1990_139_84} were proposed to describe covalent
chemical bonds in molecules when the mutual attractive forces quickly decrease
with increase of the internuclear distance. The exponential dependences of MO
and EMO potential functions from the internuclear separations describe this
type of chemical interaction well. Contrary to this case, the van der Waals
interaction in molecules is described by a potential function whose potential
well is usually significantly smaller and its depth reduces significantly
wider in comparison with those of covalent bonding. The main asymptotic term
with respect to the internuclear distance ($R$) of such a potential function is
proportional to $1/R^6$. Therefore, the EMO potential function better describes
covalent bonding rather than van der Waals bonding.

Note, that in the article
\cite{PatkowskiSpirkoSzalewicz_Science_2009_324_1548}, published shortly after
the experimental work \cite{MerittBondybeyHeaven_Science_2009_324_1548} on
Be$_2$, the authors noted that the problem with the assignment of the twelfth
vibration energy level with $\nu$=11 could have arisen due to incorrect
asymptotic behavior of the EMO potential function as internuclear distance
increases beyond 5 \AA. However, in publications
\cite{Mitin_IntJQuantumChem_2011_111_2560, Mitin_CPL_2017_682_30} the argument
has been made that the transformation of covalent bonding to the van der Waals
one begins at a distance of about 3.3 \AA. The results of \textit{ab initio}
calculations presented in work
\cite{PatkowskiSpirkoSzalewicz_Science_2009_324_1548} support this conclusion.
The \textit{ab initio} calculations of the Be$_2$ interaction potential were
performed for 20 internuclear distances and the obtained points were fitted
using a function with correct asymptotic behavior (eq. (1) in
\cite{PatkowskiSpirkoSzalewicz_Science_2009_324_1548}). As a result, the twelve
vibrational energy levels of this potential reproduce the experimental ones
with an RMS error of 3.4 cm$^{-1}$. Comparing the calculated vibrational energy
levels to the experimental values, presented in
\cite{PatkowskiSpirkoSzalewicz_Science_2009_324_1548} shows that the deviations
increase as one moves from the low-lying vibrational levels to the top ones.
Such behavior of the deviations is, once again, reasonably explained by the
fact that the basis set used in \textit{ab initio} calculations was primarily
developed for the description of covalent bonding and does not give an
equivalent description of the van der Waals interaction at large internuclear
separations. In this article, the semi-empirical ``morphed'' potentials has
also been presented, which reproduces the experimental vibrational energy
levels with an RMS error of less than 0.1 cm$^{-1}$.

\section{ The potential energy curves and its extension on a large interval}
\label{section4}

In quantum chemical calculations, the potential energy curves (PECs) of
interatomic interaction are presented in the form of numerical tables
calculated with limited accuracy and defined on a nonuniform mesh of nodes in a
finite domain of interatomic distance variation. However, for a number of
diatomic molecules the asymptotic expressions for the PEC are calculated
analytically for sufficiently large distances between the atoms
\cite{Simrnov1992, PatilTang2000, Jun-Jiang_J-MitroyADNDT2015_101_158}.

To formulate the boundary value problem (BVP) on a semiaxis, the PEC should be
continued beyond the finite interval using  additional information about the
interaction of atoms comprising the diatomic molecule at large interatomic
distances. The dominant term of the PEC at large distances is given by the van
der Waals interaction, inversely proportional to the sixth power of the
independent variable with the constant, determined from theory
\cite{Porsev_DereviankoJETP2006_102_195,SKL_Tang_PRA_2013_88_022517}.

Proceeding in this way we faced a problem how to match smoothly the PEC
asymptotic expansion with its tabulated numerical values (within the accuracy
of their calculation) at a suitable sufficiently large distance and calculate
correctly the required sets of bound, metastable and scattering states. In Ref. 
\cite{DerbovJQSRT2021_262_107529}, this problem was studied and a procedure was 
developed  for approximating MEMO PEC \cite{Mitin_CPL_2017_682_30} by Lagrange 
interpolation polynomials (LIPs). The  extension over a large interval was 
provided by a procedure of smooth matching using Hermite interpolation 
polynomials (HIPs) to preserve the PEC derivative continuity at the matching 
point \cite{GusevLNCS2014_8660_138, spie19, SPIE2020}. Such PEC construction 
was referred to as MEMO*. Below we briefly describe the basic ideas of Ref. 
\cite{DerbovJQSRT2021_262_107529}.

To describe  the Be{$_2$} diatomic molecule in the adiabatic approximation (in 
which the diagonal nonadiabatic correction is not taken into account), commonly 
referred to as Born--Oppenheimer (BO) approximation, in 
\cite{DerbovJQSRT2021_262_107529} the Schr\"odinger equation was used in the
form
\begin{eqnarray}\label{neweqold}   &&
	\left(- s_2\frac{1}  {r^2}\frac{d}
	{d r}{ r^2}\frac{d}  {d r}
	+ {V}_J( r)- E\right) \Phi_{J}(r)=0, \\
	&& V_J( r)={V}(r)+ s_2\frac{J(J+1)}{r^2},\quad
	s_2=\frac{\hbar^2}{2m}\frac{1}{\mbox{\AA}^2}.
	\nonumber
\end{eqnarray}
Here $J$ is the total angular momentum quantum number, $m=M/2=4.506$ Da is the
reduced mass of beryllium molecule expressed in  1 Da = 931.494061 MeV/c$^2$
atomic mass unit (u) \cite{nist}, 1 eV = 8065.54429 cm$^{-1}$,
$\hbar c=1973.269718$ eV \AA, and $\hbar^2/(2m)=3.741151852\cdot 10^{-8}$~\AA.
The factor $1/\mbox{\AA}^2$ in $s_2$ means that the distance $r$ between atoms
is expressed in \AA, and $s_2=3.741151852$  cm$^{-1}$. Also $E$ is the energy
in  cm$^{-1}$ and $V(r)$ is PEC in cm$^{-1}$.

For the numerical calculations, the potential, energy, and wave number in
angstroms were used
\begin{eqnarray}
	U(r)=\frac{1}{s_{2}}V(r)\cdot \mbox{\AA}^{-2}, \quad  {\cal \bar
		E}={\cal E}\cdot \mbox{\AA}^{-2},
	\quad  \bar k=\sqrt{{\cal \bar E}}=\sqrt{\cal E}\cdot \mbox{\AA}^{-1},
\end{eqnarray}
where ${\cal E}=E/s_2$, and $k=\sqrt{{\cal E}}$ are the dimensionless energy
and wave number.

\begin{figure}[t]
	\includegraphics[width=0.45\textwidth]{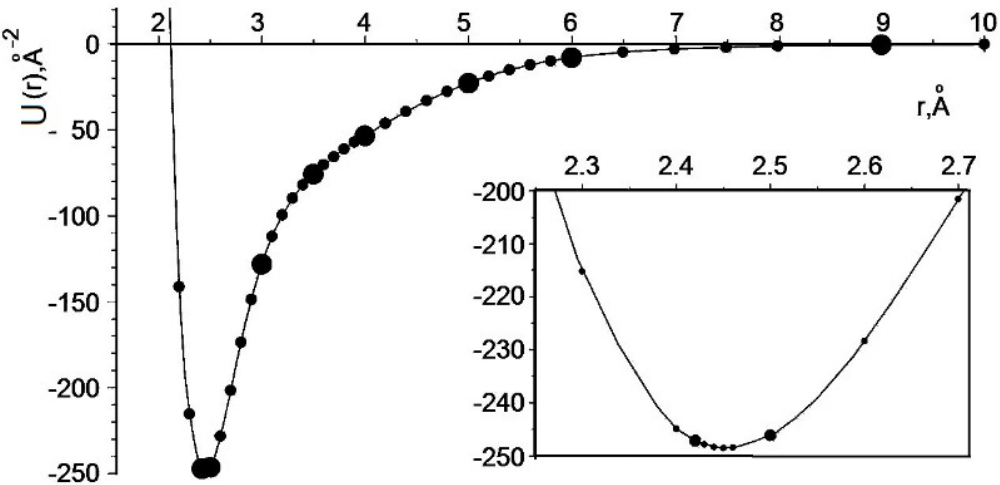}(a)
	\includegraphics[width=0.45\textwidth]{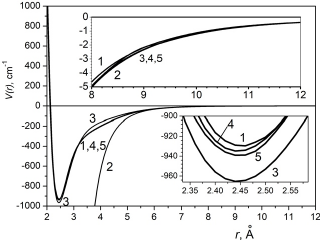}(b)
	\caption{(a) Potential $U(r)$ (\AA$^{-2}$) of the beryllium dimer as a
		function of internuclear distance $r$ (\AA) obtained by interpolating
		the	MEMO tabulated values~\cite{DerbovJQSRT2021_262_107529} (points in
		the subintervals, the boundaries of which are shown by larger dots) by
		the fifth-order	LIPs. (b) MEMO* potential $V(r)$ (points in (a) and
		line	1, Ref.~\cite{DerbovJQSRT2021_262_107529}), the asymptotic
		expansion $V_{\rm	as}(r)$ of MEMO function (line 2, Ref.
		\cite{Porsev_DereviankoJETP2006_102_195}), the analytical forms of the
		potential function $V_{an}(r)$ (line 3, Ref.
		\cite{SKL_Tang_PRA_2013_88_022517},	line 4, Ref.
		\cite{LPBMM_JChemTheorComp_2019_15_2470}, and line 5, Ref.
		\cite{MSHHLR_JCP_2014_140_064315}). $r$ is given in \AA, $V(r)$ in
		cm$^{-1}$.}
	\label{mp}
\end{figure}
\begin{figure}[t]
	\centerline{\includegraphics[width=0.8\textwidth]{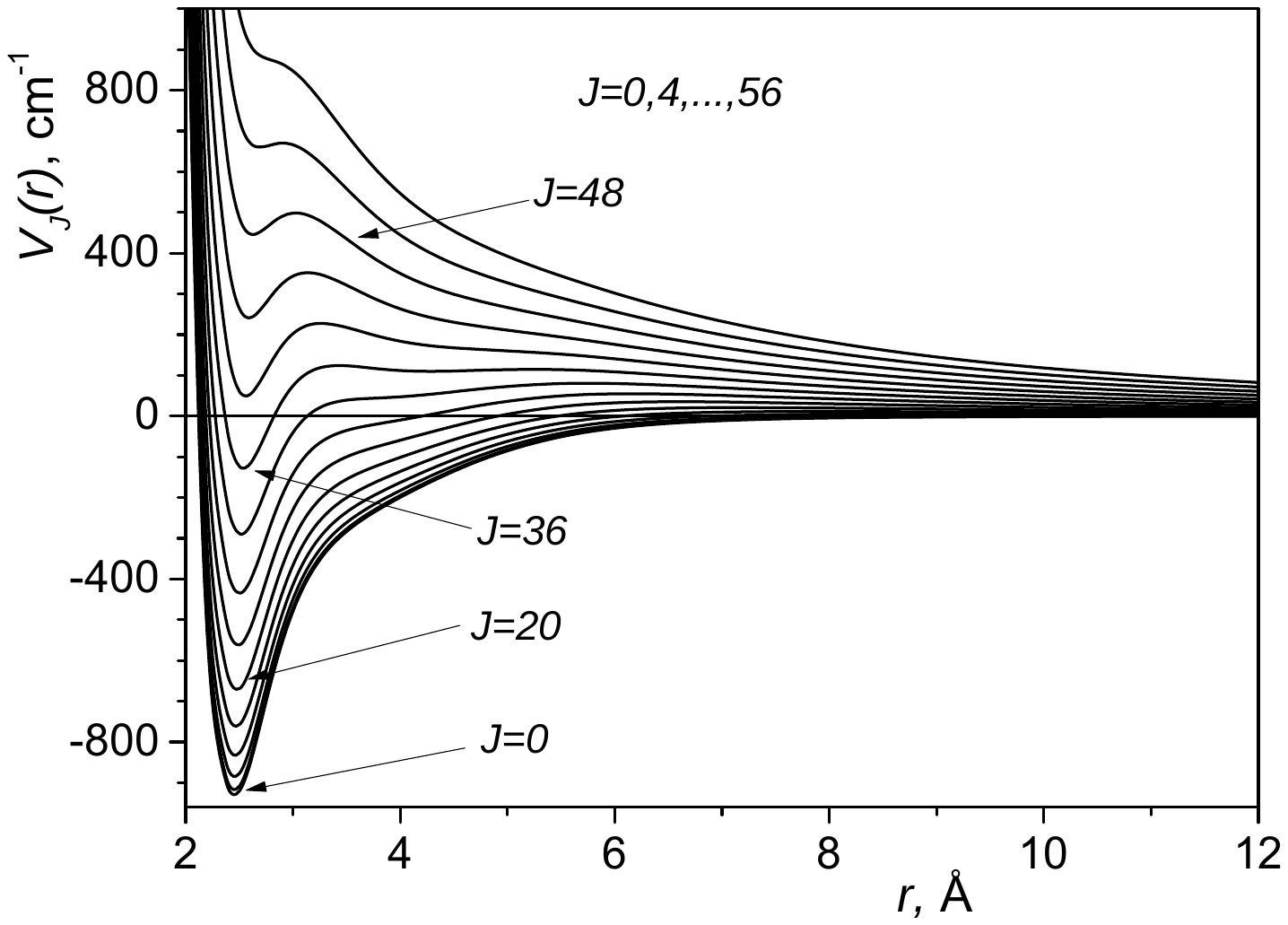}}
	\caption{MEMO* potential functions $V_J(r)$ at $J=0,4,8,\ldots,56$.}
	\label{l49}
\end{figure}

The potential $V(r)$ (in cm$^{-1}$) is given by the MEMO* potential function 
which is an approximation of the MEMO tabular values $\{V(r_i)\}_{i=1}^{76}$ in 
interval $r\in [r_1=1.5,r_{76}=48]$. Here and below, the value $r$ is given in 
units of {\AA} unless otherwise is specified. These tabular values were chosen 
to provide better approximation of the potential $V(r)$ by the fifth-order 
Lagrange interpolation polynomials (LIPs) of the variable $r$ in subintervals 
$r\in [r_{5k-4}, r_{5k+1}]$, $k=1,\ldots,15$. Indeed, one can see that Figure 
\ref{mp} displays a smooth approximation till $r_{49}=12$, where the 
approximate PEC is matched with the asymptotic potential $V^{\rm BO}_{\rm 
as}(r)=s_2U^{\rm BO}_{\rm as}(r)$ given analytically by the expansions 
\cite{Porsev_DereviankoJETP2006_102_195}
\begin{equation}\label{dde}
	U^{\rm BO}_{\rm as}(r)=s_1 \tilde V^{\rm BO}_{\rm as}(r),\quad\tilde
	V_{\rm as}^{\rm BO}(r)=-\left(214
	Z^{-6}+{10230}Z^{-8}+{504300}{Z^{-10}}\right),
\end{equation}
where $s_1= \mbox{aue} /s_{2}=58664.99239$ or $s_1s_2=\mbox{aue}=219474.6314$
cm$^{-1}$, $Z=r/s_3$ and $s_3=0.52917$ is the Bohr radius in \AA. This fact
allowed considering the interval $r\in[r_{\rm match}\geq12,\infty)$ as possible
for using the asymptotic  potential $V^{\rm BO}_{\rm as}(r)$ at large $r$ and
executing conventional calculations based on tabular values of $V(r)$ in the
finite interval $r\in [r_{1}, r=12]$ (see also
\cite{LPBMM_JChemTheorComp_2019_15_2470}).

The MEMO tabular values for $r\in\{{r_{42}}=6.5,...,r_{48}=11\}$ are smaller 
than the asymptotic ones by $5.5\div6$\%, for $r=r_{51}=14$ exceed the 
asymptotic ones by 8\%, and beyond the interval 
$r\in[r_{41}=6.0,\ldots,r_{52}=15]$ the difference is more than 10\%.

Based on this fact, the potential $V(r)$ in subintervals $r\in [r_{5k-4},
r_{5k+1}]$, $k=1,\ldots,9$ was approximated by the fifth-order Lagrange
interpolation polynomials (LIPs) of the variable $r$ in the interval
$r\in [r_1,r_{46}=9]$. In subinterval $r\in[r_{46}=9,r_{\rm match}=14]$ 
the potential $V(r)$ was approximated by the fourth-order HIPs using the values 
of the potential $V(r)$ at the points $r=\{r_{46}=9,r_{47}=10,r_{48}=11\}$ and 
the values of the asymptotic potential $V_{\rm as}(r)$ and its derivative $d 
V_{\rm as}(r)/dr$ at the point $r=r_{\rm match}=14$. In the interval 
$r\in[r_{\rm match}=14,\infty)$, the potential $V(r)$ is approximated by the 
asymptotic expansion (\ref{dde}). Let us emphasize that this approximation 
using the HIP provides a smooth matching of the interpolated values of the 
tabulated function with its asymptotic continuation, in particular case 
(\ref{dde}),  in comparison with the conventional approximation by LIPs 
providing no continuity of the derivative at the boundary of joined intervals 
\cite{GusevLNCS2014_8660_138}. Potentials $V_J(r)$ are displayed in Figure 
\ref{l49} at $J=0,\ldots,56$ with the step 4.

In Ref. \cite{LPBMM_JChemTheorComp_2019_15_2470}, the potential $ V(r)$ (in 	
cm$^{-1}$) (see Figure \ref{mp}) is given by the BO potential function plus
relativistic potential function marked as STO with tabular values
$\{V(Z_i)\}_{i=1}^{28}$ in the interval $Z\in [Z_1=3.75,Z_{28}=25]$ a.u. which
corresponds to $r\in [r_1=1.9843,r_{28}=13.229]$. One can see that these
tabular values were chosen to provide the best approximation of the potential
$V(r)$ by the fourth-order LIPs of the variable $r$ in subintervals
$r\in[r_{4k-3}, r_{4k+1}]$, $k=1,\ldots,6$. On interval $Z\in[Z_{25},Z_{\rm
match}=27.5]$ a.u. the potential $V(r)$ is approximated by the fifth-order HIP 
using the  values of the potential $V(Z)$ at the points 
$Z=\{Z_{25}=17.5,Z_{26}=20.0,Z_{27}=22.5,Z_{28}=25.0\}$ a.u. and the values of
the asymptotic potential $V_{\rm as}(r)$ and its derivative	$d V_{\rm
as}(Z)/dZ=s_3d V_{\rm as}(r)/dr$ at the point $Z=Z_{\rm match}=27.5$ a.u. In
the interval $r\in[r_{\rm match}=14.552,\infty)$ the potential
$V_{\rm as}(r)=s_2 U_{\rm as}(r)$ is approximated by the asymptotic expansion
\begin{eqnarray}\label{ddf} &&
	U_{\rm as}(r)=s_1 \tilde V_{\rm as}(r),\quad\tilde V_{\rm
		as}(r)=\tilde V_{\rm as}^{\rm BO}(r)+\tilde V_{\rm as}^{\rm rel}(r),
	\\ && \tilde V_{\rm as}^{\rm rel}(r)=-\left(1.839 \cdot10^{-4}
	Z^{-4}+0.11944 Z^{-6}+19.582Z^{-8}-1323.5{Z^{-10}}\right),
	\nonumber
\end{eqnarray}
where $\tilde V_{\rm as}^{\rm BO}(r)$ is given by Eq. (\ref{dde}) and $\tilde 
V_{\rm as}^{\rm rel}(r)$ is taken from Ref. 
\cite{LPBMM_JChemTheorComp_2019_15_2470}. The STO PEC constructed in such a way 
was called STO*. Note that using a similar behavior of MEMO and STO potential 
functions on the interval $r\in [12,14]$ one can use also Eq. (\ref{ddf}) for 
matching MEMO potential in interval $r\in [14,\infty)$, because  it has been 
calculated taken into account the relativistic effects 
\cite{DouglasKroll_AnnPhys_1974_82_89, Hess_PRA_1985_32_756, 
Hess_PRA_1986_33_3742}.

For comparison we show in Figure \ref{mp}b the  potential function $V(r)$, its
asymptotic expansion $V_{\rm as}(r)$ and the analytical potential functions
$V_{an}(r)$ in a.u. (converted into cm$^{-1}$), proposed in Ref.
\cite{SKL_Tang_PRA_2013_88_022517}. The approximated MEMO* potential
function $V(r)$ has a minimum $-D_{e}({\rm MEMO^*})=V(r_e)=-929.804$
cm$^{-1}$ at the equilibrium point $r_e=2.4534$ \AA, which is higher than the
analytical potential function $V_{an}(r)$ in the vicinity of this point,
$-D_e({\rm Sheng})=V_{\rm an}(r_e)=-948.3$ cm$^{-1}$. On the contrary, in the
interval $r \in (3.2,6.1)$ the analytical potential function $V_{\rm an}(r)$ is
greater than $V(r)$. For $r \in (2.3, 12)$, the MEMO potential slightly exceeds
the STO one, which, in turn, is a bit higher than the MLR and CPE potentials.
Thus, using the accepted approximations we have the MEMO and STO potential
functions $V(r)$ approximated in the analytical form in interval $r\in(1.9,14)$
and its smooth continuation at $r\geq14$ by means of the asymptotic expressions
(\ref{dde}) and (\ref{ddf}).

\begin{table}[t]
	\caption{Vibrational-rotational bound states $-E_{\nu J}$ (in cm$^{-1}$)
		of the beryllium dimer. For each $J$ in upper line MEMO* and in	lower 
		line STO* with relativistic corrections. Adopted from 
		\cite{DerbovJQSRT2021_262_107529}.}
	\label{ls}
	\begin{tabular}{|r|rrrrr|rrrrr|rr|}
		\hline
		$J$&$v=0$&1&2&3&4&5&6&7&8&9&10&11\\ \hline
		0& 806.0& 583.5& 408.7& 288.3& 211.1& 154.1& 107.1&  68.3&  37.8&
		16.3& 4.4&    0.3\\
		0& 807.7& 584.1& 409.4& 288.3& 211.9& 155.4& 108.3&  69.5& 	39.0&
		17.5& 5.0&    0.4\\ \hline
		1& 804.8& 582.4& 407.7& 287.5& 210.4& 153.5& 106.6&  67.8&  37.4&
		16.0& 4.2&    0.2\\
		1& 806.5& 583.0& 408.4& 287.5& 211.2& 154.8& 107.7&  69.0&
		38.6&  17.2& 4.8&    0.3\\ \hline
		2& 802.4& 580.1& 405.7& 285.8& 209.0& 152.3& 105.5&  66.9&  36.6&
		15.4& 3.8&    0.1\\
		2& 804.0& 580.8& 406.4& 285.8& 209.8& 153.5& 106.7&  68.1&  37.8&
		16.6& 4.4&    0.1\\ \hline
		3& 798.7& 576.7& 402.6& 283.2& 206.9& 150.4& 103.8&  65.5&  35.4&
		14.5& 3.2&       \\
		3& 800.4& 577.4& 403.5& 283.3& 207.7& 151.7& 105.1&  66.7&  36.7&
		15.7& 3.8&       \\ \hline
		4& 793.9& 572.2& 398.6& 279.8& 204.1& 148.0& 101.7&  63.6&  33.8&
		13.3& 2.4&       \\
		4& 795.5& 573.0& 399.5& 279.9& 204.9& 149.3& 102.9&  64.8&  35.1&
		14.5& 3.0&       \\ \hline
		5& 787.8& 566.5& 393.6& 275.6& 200.5& 144.9&  99.0&  61.2&  31.8&
		11.8& 1.5&       \\
		5& 789.5& 567.4& 394.5& 275.8& 201.5& 146.3& 100.2&  62.5&  33.2&
		13.0& 2.1&       \\ \hline
		6& 780.5& 559.7& 387.5& 270.6& 196.3& 141.2&  95.7&  58.4&  29.5&
		10.0& 0.5&       \\
		6& 782.2& 560.7& 388.6& 270.8& 197.3& 142.6&  97.0&  59.7&  30.9&
		11.2& 1.0&       \\ \hline
		7& 772.1& 551.8& 380.5& 264.7& 191.4& 137.0&  92.0&  55.2&  26.8&
		8.1&    &       \\
		7& 773.7& 552.9& 381.7& 265.0& 192.5& 138.4&  93.3&  56.5&  28.2&
		9.2&    &       \\ \hline
		8& 762.4& 542.8& 372.5& 258.0& 185.9& 132.1&  87.7&  51.5&  23.8&
		5.9&    &       \\
		8& 764.1& 544.0& 373.8& 258.4& 187.0& 133.6&  89.1&  52.9&  25.2&
		7.0&    &       \\ \hline
		9& 751.5& 532.7& 363.5& 250.5& 179.7& 126.7&  83.0&  47.4&  20.5&
		3.5&    &       \\
		9& 753.2& 534.0& 364.9& 251.0& 180.9& 128.3&  84.4&  48.8&  21.9&
		4.6&    &       \\ \hline
		10& 739.4& 521.4& 353.6& 242.2& 172.8& 120.7&  77.7&  42.9& 16.8&
		1.0&    &       \\
		10& 741.1& 522.9& 355.1& 242.8& 174.2& 122.4&  79.2&  44.3& 18.3&
		2.0&    &       \\ \hline
	\end{tabular}
\end{table}
\begin{table}[t]
	\caption{Continuation of Table \ref{ls}.}
	\label{lsss}
	\begin{tabular}{|r|rrrrr|rrrr|}
		\hline
		$J$&$\nu=0$&1&2&3&4&5&6&7&8\\ \hline
		11& 726.2& 509.1& 342.6& 233.2& 165.3& 114.2&  72.0&  38.0&12.9\\
		11& 727.9& 510.7& 344.4& 233.9& 166.8& 115.9&  73.6&  39.5&14.4\\ \hline
		12& 711.7& 495.6& 330.8& 223.4& 157.2& 107.1&  65.8&  32.8& 8.7\\
		12& 713.4& 497.4& 332.7& 224.3& 158.8& 108.9&  67.4&  34.3&10.2\\ \hline
		13& 696.0& 481.1& 318.0& 212.9& 148.5&  99.5&  59.2&  27.1& 4.4\\
		13& 697.8& 483.1& 320.1& 213.9& 150.2& 101.4&  60.9&  28.7& 5.9\\ \hline
		14& 679.2& 465.5& 304.3& 201.7& 139.2&  91.4&  52.2&  21.2& $-$\\
		14& 681.0& 467.7& 306.6& 202.8& 141.0&  93.4&  53.9&  22.9& 1.4\\ \hline
		15& 661.2& 448.8& 289.7& 189.8& 129.3&  82.8&  44.7&  15.0&    \\
		15& 663.0& 451.2& 292.2& 191.1& 131.3&  84.8&  46.5&  16.7&    \\ \hline
		16& 642.1& 431.0& 274.2& 177.3& 118.9&  73.8&  36.9&   8.5&    \\
		16& 643.8& 433.7& 276.9& 178.7& 121.1&  75.9&  38.8&  10.3&    \\ \hline
		17& 621.8& 412.2& 257.9& 164.2& 108.0&  64.3&  28.7&   1.9&    \\
		17& 623.5& 415.1& 260.8& 165.8& 110.3&  66.5&  30.7&   3.7&    \\ \hline
		18& 600.3& 392.4& 240.7& 150.4&  96.6&  54.4&  20.3&      &    \\
		18& 602.1& 395.5& 243.8& 152.2&  99.0&  56.6&  22.3&      &    \\ \hline
		19& 577.7& 371.5& 222.8& 136.2&  84.8&  44.1&  11.5&      &    \\
		19& 579.4& 374.9& 226.0& 138.2&  87.3&  46.4&  13.6&      &    \\ \hline
		20& 553.9& 349.7& 204.0& 121.5&  72.5&  33.5&   2.6&      &    \\
		20& 555.7& 353.3& 207.5& 123.7&  75.2&  35.9&   4.8&      &    \\ \hline
	\end{tabular}
\end{table}
\begin{table}[t]
	\caption{Continuation of Table \ref{ls}.}
	\label{lss}
	\begin{tabular}{|r|rrrrrr||r|rrr|}
		\hline
		$J$&$\nu=0$&1&2&3&4&5&$J$&$\nu=0$&1&2\\ \hline
		21& 529.1& 326.8& 184.6& 106.3&  59.8&  22.5&29& 290.8& 111.0 &10.7  \\
		21& 530.8& 330.7& 188.2& 108.7&  62.6&  25.0&29& 292.7& 117.0 &14.9  \\
		\hline
		22& 503.1& 303.0& 164.4&  90.8&  46.8&  11.4&30& 256.3&  80.4 &      \\
		22& 504.9& 307.2& 168.2&  93.4&  49.7&  13.9&30& 258.2&  86.6 &      \\
		\hline
		23& 476.0& 278.2& 143.6&  74.9&  33.4&  0.01&31& 220.8&  49.1 &      \\
		23& 477.8& 282.7& 147.6&  77.8&  36.4&   2.5&31& 222.7&  55.6 &      \\
		\hline
		24& 447.8& 252.5& 122.3&  58.8&  19.8&      &32& 184.4&  17.2 &      \\
		24& 449.6& 257.2& 126.4&  61.9&  22.9&      &32& 186.3&  23.9 &      \\
		\hline
		25& 418.5& 225.9& 100.4&  42.5&   5.9&      &33& 146.9&       &      \\
		25& 420.3& 230.9& 104.7&  45.8&   9.1&      &33& 148.9&       &      \\
		\hline
		26& 388.2& 198.4&  78.2&  26.0&      &      &34& 108.6&       &      \\
		26& 390.0& 203.7&  82.5&  29.6&      &      &34& 110.6&       &      \\
		\hline
		27& 356.7& 170.1&  55.7&   9.5&      &      &35&  69.3&       &      \\
		27& 358.6& 175.6&  60.1&  13.3&      &      &35&  71.4&       &      \\
		\hline
		28& 324.3& 140.9&  33.1&      &      &      &36&  29.2&       &      \\
		28& 326.1& 146.7&  37.5&      &      &      &36&  31.3&       &      \\
		\hline
	\end{tabular}
\end{table}

The MAPLE and FORTRAN programs used to get the analytical form of approximation
for the MEMO \cite{Mitin_CPL_2017_682_30} and STO
\cite{LPBMM_JChemTheorComp_2019_15_2470} potential functions $V_J(r)$,		
respectively, extended over large intervals of internuclear distance $r$ with
the help of asymptotic expressions (\ref{dde}) and (\ref{ddf}), are given in
the supplementary material of Refs. \cite{DerbovJQSRT2021_262_107529,	
Supplementary}. Below we use the notation MEMO* and STO* for these potential 
functions.

\section{Bound states of the beryllium dimer} \label{bsbd}

The vibrational-rotational spectrum of the real-valued eigenenergies $E_{\nu 
J}$ and the corresponding eigenfunctions $\Phi_{\nu J}(r)$ of the bound states 
of the BVP for Eq. (\ref{neweqold}) were calculated 
\cite{spie19,SPIE2020,DerbovJQSRT2021_262_107529} using the FEM programs KANTBP 
4M~\cite{KANBP4M} and KANTBP 3.0 \cite{kantbp3.0} on the finite element mesh
$\Omega_{1}(r)=\{$1.9(0.1)2.4(0.05)2.8(0.1)4.0(0.2)5.0(0.5)8(2) 20(5)40$\}$,
where the number in parentheses $(x)$ is the size of subinterval, with the
second-type or Neumann boundary conditions (BCs) on the boundary points of the
mesh. In the BVP solution at all finite elements of the mesh the local functions
were represented by fifth-order HIPs.

Table \ref{Table_A4} (columns MEMO* and STO*) presents the results of using FEM
programs KANTBP 4M and KANTBP 3.0 to calculate 12 energy eigenvalues of the
beryllium dimer. It shows the eigenvalues calculated with the MEMO potential
function \cite{Mitin_CPL_2017_682_30} and the corresponding approximation MEMO*
from \cite{DerbovJQSRT2021_262_107529} and the previous Section \ref{section4}.
In contrast to the original EMO function, which was used to describe the
experimental (Exp) vibrational levels
\cite{MerittBondybeyHeaven_Science_2009_324_1548}, it has not only the correct
dissociation energy, but also describes all twelve vibrational energy levels
with the RMS error less than 0.4 cm$^{-1}$.	The Table \ref{Table_A4} also shows
the results of direct-potential-fit analysis using the MLR and CPE functions
alongside with the EMO potential function \cite{MSHHLR_JCP_2014_140_064315},
and CV+F+R potential function \cite{Koput_PCCP_2011_13_20311} discussed early in
Section \ref{section2}. Similar results FCI/CBS+corr were obtained by Lesiuk
et. al. \cite{LPBMM_JChemTheorComp_2019_15_2470} and STO* Derbov et. al.
\cite{DerbovJQSRT2021_262_107529}. Their PEC lie below the MEMO one and also
include the correct long-range behavior displayed in Figure \ref{mp}. As a
consequence, one can see from the Tables \ref{ls}--\ref{lss}, that the
corresponding results provide the theoretical {\it lower estimates} whereas
MEMO* and MEMO results give the {\it upper estimates} for the discrete spectrum
of the beryllium dimer at both $J=0$ and $J>0$ in accordance with
\cite{CourantHilbert1989}. One can see also that the STO* eigenenergies
calculated using the STO* approximation of the tabulated PEC STO give smaller
RMS error  0.7 cm$^{-1}$ in comparison with RMS error 1.0 cm$^{-1}$ of the
FCI/CBS+corr eigenenergies calculated using the analytical fit of STO
PEC\cite{LPBMM_JChemTheorComp_2019_15_2470}.

\begin{figure}[t]
	\centerline{\includegraphics[width=0.45\textwidth]{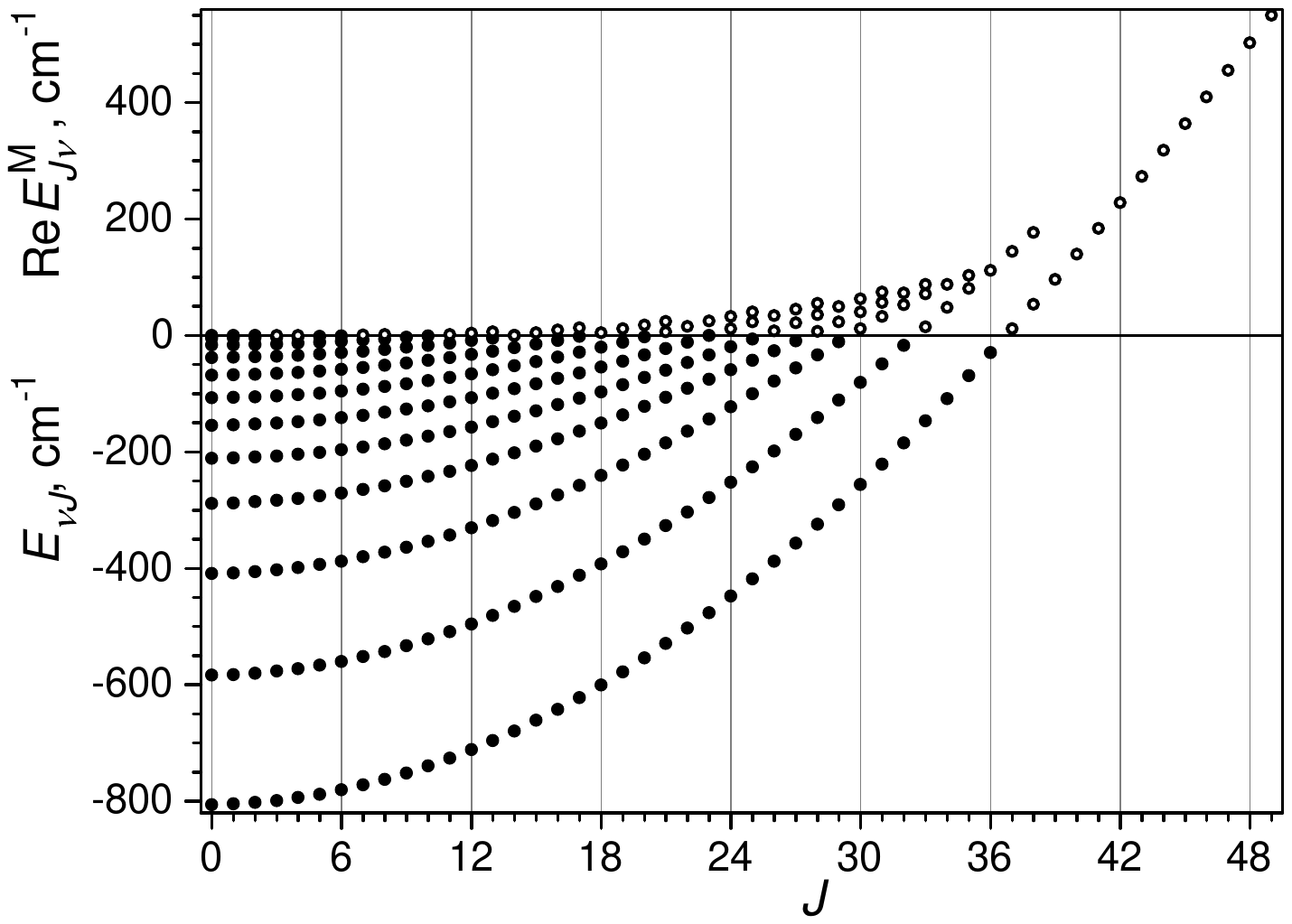} a)
	            \includegraphics[width=0.45\textwidth]{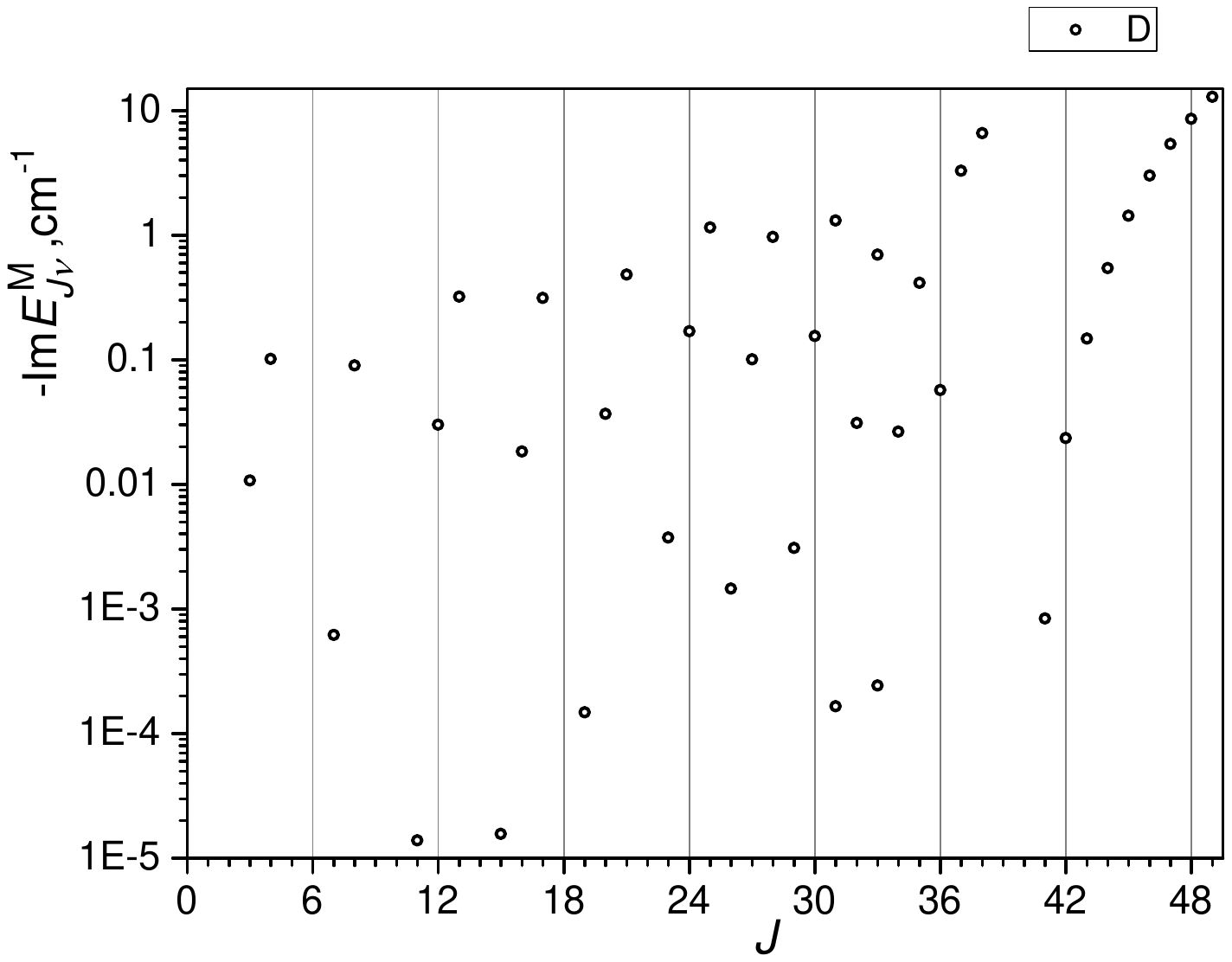} b)}
	\caption{a) Eigenenergies $E_{\nu J}$ of vibrational-rotational bound
	states (black dots), the real part $\Re E_{J\nu}^{M}$ (empty circles) and
	b) minus the imaginary part $-\Im E_{J\nu}^{M}$ of complex eigenenergies
	$E_{J\nu}^{M}=\Re E_{J\nu}^{M}+\imath\Im E_{J\nu}^{M}$ of
	rotational-vibrational metastable states in	cm$^{-1}$.}
	\label{l49b}
\end{figure}

As shown in Refs. \cite{spie19,SPIE2020,DerbovJQSRT2021_262_107529}, the 
potential functions MEMO* and STO* $V_J(r)$ from $J=0$ to $J=36$ support
252 and 253 vibrational-rotational energy levels $E_{\nu J}$, respectively,
presented in Tables \ref{ls}--\ref{lss}. Figure \ref{l49b}a) (black dots) shows
also the vibrational-rotational spectrum $E_{\nu J}$ (in cm$^{-1}$) of Be$_2$
vs $J$ for MEMO* PEC. One can see that these potential functions $V_J(r)$ at
$J=0$, $J=1$ and $J=2$ support 12 vibrational energy levels and there is no
energy level $E_{\nu=8,J=14}$ for MEMO* PEC.

Here and below the results presented in figures are calculated with MEMO* PEC.

\begin{table}
	\caption{The vibrational-rotational metastable states $E_{J\nu}^{M}=\Re
	{E}_{J\nu}^{M}+\imath\Im { E}_{J\nu}^{M}$ (in cm$^{-1}$) of Be$_2$, where
	``eps'' means that $-10^{-5}<\Im { E}_{J\nu}^{M}<0$ (in cm$^{-1}$). From
	left to right MEMO* and STO* with relativistic corrections. $V_J^{\min}$
	and $V_J^{\max}$ are minimal and maximum values of potentials $V_J(r)$ (in
	cm$^{-1}$) at different values $J$ of the total angular	momentum. The
	metastable states with real part of energy $\Re E_{J\nu}^{M}$ greater then
	the top of potential barrier $V_J^{\max}$ are marked by	asterisk. Adopted 
	from \cite{DerbovJQSRT2021_262_107529}.}
	\label{ms1}
	\begin{tabular}{|rr|rr|rr|rr|rr|}
		\hline
		$J$&$\nu$& \multicolumn2{|c|}{MEMO*}& \multicolumn2{|c|}{STO*}&
		\multicolumn2{|c|}{MEMO*}& \multicolumn2{|c|}{STO*} \\
		\hline
		&  &$V_J^{\min}$&$V_J^{\max}$&$V_J^{\min}$&$V_J^{\max}$&$\Re E$&$-\Im
		E$&$\Re E$&$-\Im E$\\
		\hline
		0&  &-929.80&  0.00&-934.39& 0.00&        &      &        &      \\
		1&  &-928.55&  0.01&-933.15& 0.01&        &      &        &      \\
		2&  &-926.07&  0.04&-930.66& 0.04&        &      &        &      \\
		3&11&-922.34&  0.12&-926.92& 0.11&   0.081&0.010 &  & \\
		{*}4&11&-917.37&  0.23&-921.93& 0.24&   0.281&0.101  &0.256   &
		0.063    \\
		5&&-911.16&  0.42&-915.70& 0.43&   & &  & \\
		6&&-903.71&  0.71&-908.22& 0.71&   & &        &      \\
		7&10&-895.03&  1.11&-899.59& 1.09&   0.503&6.1e-4&   0.118&eps\\
		8&10&-885.11&  1.62&-889.52& 1.57&   1.509&0.090 &   1.254&0.026 \\
		{*}9&10&-873.95&  2.24&-878.41& 2.18&   & &   2.327&0.246 \\
		10&  &-861.57&  3.00&-866.00& 2.92&        &      &        &      \\
		11& 9&-847.96&  3.92&-852.36& 3.82&   1.554&1.3e-5   &   0.641&eps   \\
		12& 9&-833.12&  5.02&-837.49& 5.04&   4.052&0.030 &   3.308&3.8e-3 \\
		{*}13& 9&-817.06&  6.27&-821.41& 6.27&   6.356&0.321 &   5.788&0.152 \\
		\hline
		14& 8&-799.78&  7.73&-804.11& 7.64&   0.083&eps   &        &      \\
		{*}14& 9&-799.78&  7.73&-804.11& 7.64&        &      &   8.135&0.688 \\
		\hline
		15& 8&-781.28&  9.39&-785.60& 9.18&   4.605&1.5e-5 &   3.126&eps   \\
		\hline
		16& 8&-761.57& 11.26&-765.89&10.94&   8.992&0.018 &   7.651&2.2e-3\\
		17& 8&-740.65& 13.34&-744.97&12.94&  13.016&0.314 &   11.896&0.124 \\
		\hline
		18& 7&-718.53& 15.66&-722.85&15.19&   4.788&eps   &   2.916&eps   \\
		{*}18& 8&-718.53& 15.66&-722.85&15.19&        &      &   15.816&0.722\\
		\hline
		19& 7&-695.21& 18.22&-699.54&17.69&  11.517&1.4e-4&   9.635&eps   \\
		20& 7&-670.68& 21.05&-675.04&20.44&  17.991&0.036 &   16.196&8.2e-3\\
		\hline
		21& 6&-645.00& 24.16&-649.36&23.45&   6.403&eps   &   4.200&eps   \\
		21& 7&-645.00& 24.16&-649.36&23.45&  23.915&0.481 &   22.276&0.230 \\
		\hline
		22& 6&-618.13& 27.55&-622.51&26.73&  15.497&eps   &   13.264&eps   \\
		23& 6&-590.09& 31.24&-594.48&30.30&  24.444&3.7e-3&  22.218&7.5e-4\\
		\hline
	\end{tabular}
\end{table}
\begin{table}
	\caption{Continuation of Table \ref{ms1}. }
	\label{ms3}
	\begin{tabular}{|rr|rr|rr|rr|rr|}
		\hline
		$J$&$ \nu$&\multicolumn2{|c|}{MEMO*}&\multicolumn2{|c|}{STO*}&
		\multicolumn2{|c|}{MEMO*} &\multicolumn2{|c|}{STO*}\\
		\hline
		&  &$V_J^{\min}$&$V_J^{\max}$&$V_J^{\min}$&$V_J^{\max}$&$\Re E$&$-\Im
		E$&$\Re E$&$-\Im E$\\
		\hline
		24& 5&-560.92& 35.25&-565.29&34.17&  11.484&eps   &   8.853&eps   \\
		24& 6&-560.92& 35.25&-565.29&34.17&  32.872&0.168 &  30.741&0.073 \\
		\hline
		25& 5&-531.76& 39.58&-534.93&38.36&  22.998&eps   &  20.324&eps   \\
		{*}25& 6&-531.76& 39.58&-534.93&38.36&  40.609&1.155 & 38.526&0.764 \\
		\hline
		26& 4&-500.63& 44.26&-503.43&42.90&   7.996&eps   &   4.773&eps   \\
		26& 5&-500.63& 44.26&-503.43&42.90&  34.354&1.4e-3&  31.669&3.2e-4\\
		\hline
		27& 4&-468.31& 49.30&-470.77&47.80&  22.032&eps   &  18.779&eps   \\
		27& 5&-468.31& 49.30&-470.77&47.80&  45.187&0.100 &  42.578&0.044 \\
		\hline
		28& 3&-434.84& 54.73&-436.96&53.00&   6.963&eps   &   3.009&eps   \\
		28& 4&-434.84& 54.73&-436.96&53.00&  35.991&eps   &  32.731&eps   \\
		{*}28& 5&-434.78& 54.73&-436.96&53.00&  55.158&0.963 & 52.604&0.633 \\
		\hline
		29& 3&-400.28& 60.57&-402.02& 58.57&  23.517&eps   &  19.452&eps \\
		29& 4&-400.28& 60.57&-402.02& 58.57&  49.669&3.0e-4&  46.445&8.5e-4 \\
		\hline
		30& 2&-364.63& 66.91&-365.94& 64.70&  11.354&eps   &   7.180&eps  \\
		30& 3&-364.63& 66.91&-365.94& 64.70&  40.058&eps   &  35.968&eps  \\
		30& 4&-364.63& 66.91&-365.94& 64.70&  62.639&0.155 &  59.549&0.076 \\
		\hline
		31& 2&-327.89& 73.60&-328.74& 71.30&  32.621&eps   &  28.549&eps  \\
		31& 3&-327.89& 73.60&-328.74& 71.30&  56.534&1.6e-4&  52.550&3.0e-5 \\
		{*}31& 4&-327.89& 73.60&-328.74& 71.30&  74.626&1.306 & 71.589&0.939\\
		\hline
		32& 2&-290.09& 80.68&-290.41& 78.37&  52.660&eps   &  48.671&eps \\
		32& 3&-290.09& 80.68&-290.41& 78.37&  72.662&0.030 &  68.982&0.011 \\
		\hline
		33& 1&-251.24& 88.21&-250.96& 85.94&  15.028&eps   &   8.238&eps   \\
		33& 2&-251.24& 88.21&-250.96& 85.94&  71.131&2.4e-4&   67.206&7.1e-5 	
		\\
		33& 3&-251.24& 88.21&-250.96& 85.94&  87.630&0.696 &   84.364&0.476
		\\
		\hline
		34& 1&-211.35& 96.30&-210.39& 94.04&  47.644&eps   &  40.779&eps  \\
		34& 2&-211.35& 96.30&-210.39& 94.04&  87.890&0.026   &  84.078&0.012  \\
		\hline
		35& 1&-170.43&105.04&-168.75&102.73&  80.254&eps   &  73.432&eps  \\
		35& 2&-170.43&105.04&-168.75&102.73& 102.939&0.414   &  99.326&0.266 \\
		\hline
		36& 1&-128.50&124.02&-125.94&120.39& 111.593&0.057 & 105.389&6.8e-3 \\
		\hline
	\end{tabular}
\end{table}
\begin{table}
	\caption{Continuation of Table \ref{ms3}.}
	\label{ms2}
	\begin{tabular}{|rr|rr|rr|rr|rr|}
		\hline
		$J$&$\nu$&\multicolumn2{|c|}{MEMO*}&\multicolumn2{|c|}{STO*}
		&\multicolumn2{|c|}{MEMO*} &\multicolumn2{|c|}{STO*}\\
		\hline
		&  &$V_J^{\min}$&$V_J^{\max}$&$V_J^{\min}$&$V_J^{\max}$&$\Re E$&$-\Im
		E$&$\Re E$&$-\Im E$\\
		\hline
		37& 0& -85.58&147.94& -83.10&144.01&  11.780&eps   &   9.538&eps  \\
		37& 1& -85.58&147.94& -83.10&144.01& 144.261&3.272 & 137.920&2.408 \\
		\hline
		38& 0& -41.68&173.18& -38.98&168.91&  53.590&eps   & 51.233&eps  \\
		{*}38& 1& -41.68&173.18& -38.98&168.91& 177.154& 6.591&170.793&5.797\\
		\hline
		39& 0&   3.17&199.73&   6.08&195.12&  96.169&eps   & 93.672&eps \\
		40& 0&  48.96&227.57&  52.02&222.65& 139.466&eps   & 136.795&eps \\
		41& 0&  95.66&256.70&  98.82&251.52& 183.406&8.3e-4& 180.520&1.2e-3	\\
		42& 0& 143.26&287.15& 146.44&281.76& 227.880&0.023 & 224.726&0.030 \\
		43& 0& 191.72&318.91& 194.85&313.39& 272.755&0.148 & 269.266&0.175 \\
		44& 0& 241.01&352.03& 244.03&346.07& 317.922&0.544 & 314.022&0.625 \\
		45& 0& 291.11&386.52& 293.90&379.47& 363.371&1.432 & 358.983&1.622 \\
		46& 0& 341.98&422.41& 344.47&413.61& 409.201&3.008 & 404.261&3.371 \\
		{*}47& 0& 393.56&459.75& 395.60&448.50& 455.552&5.372 & 450.005&5.993 \\
		{*}48& 0& 445.83&498.57& 447.31&487.87& 502.577&8.560 &496.339&9.508 \\
		{*}49& 0& 498.71&538.92& 499.48&528.37& 550.248&12.903 &      &      \\
		\hline
	\end{tabular}
\end{table}

\begin{figure}[t]
	\includegraphics[width=0.47\textwidth]{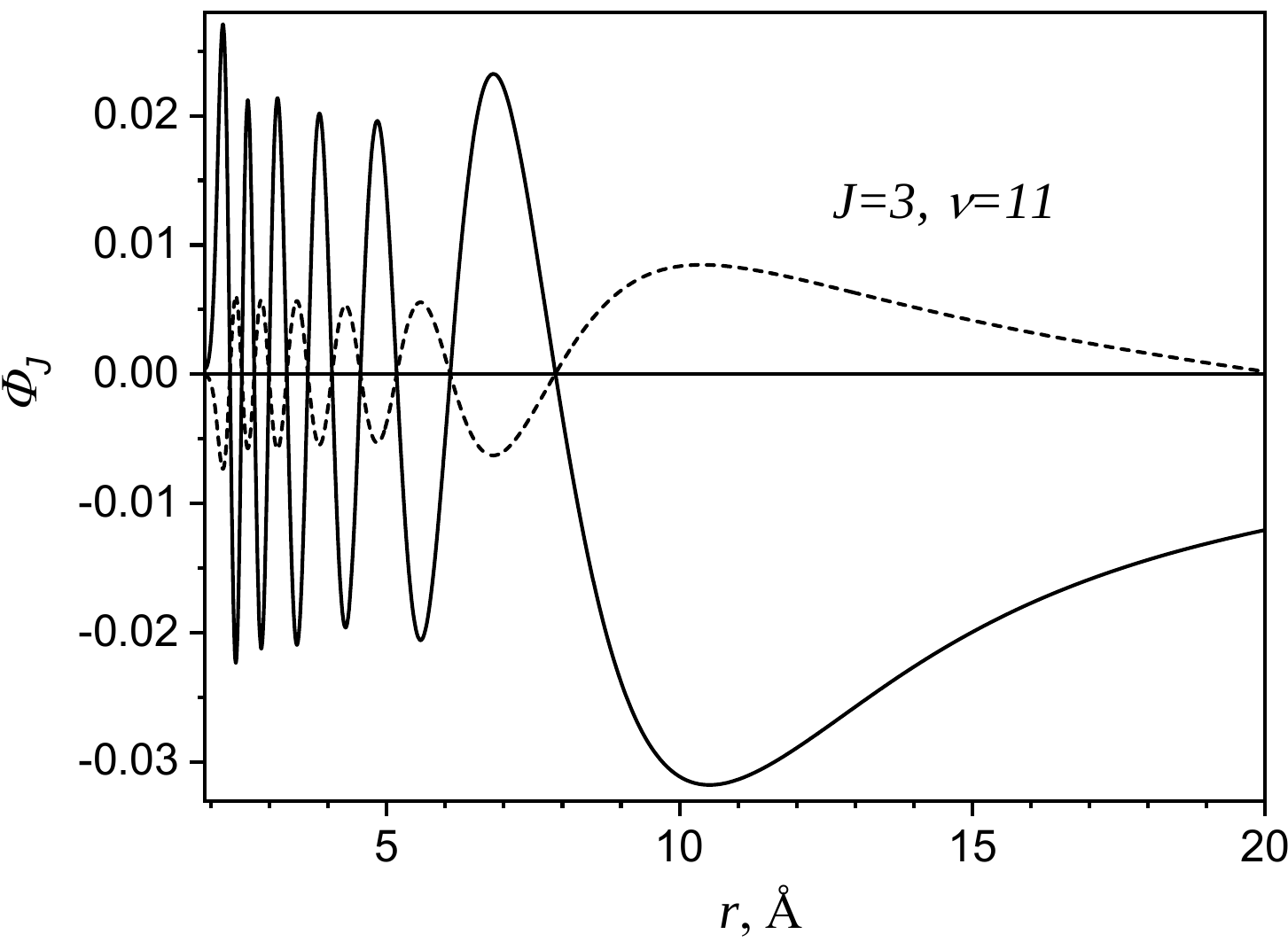}
	\includegraphics[width=0.47\textwidth]{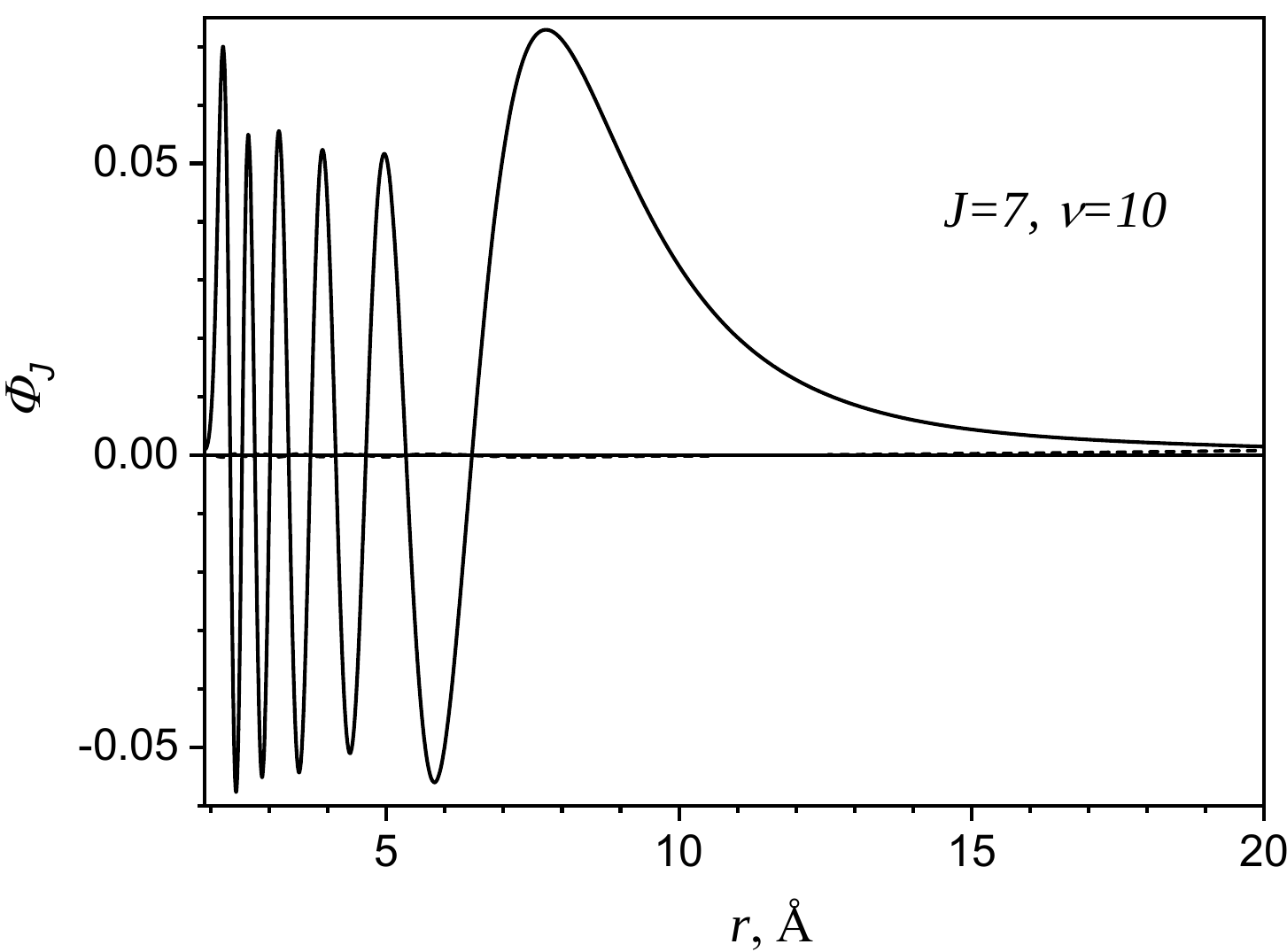}
	\includegraphics[width=0.47\textwidth]{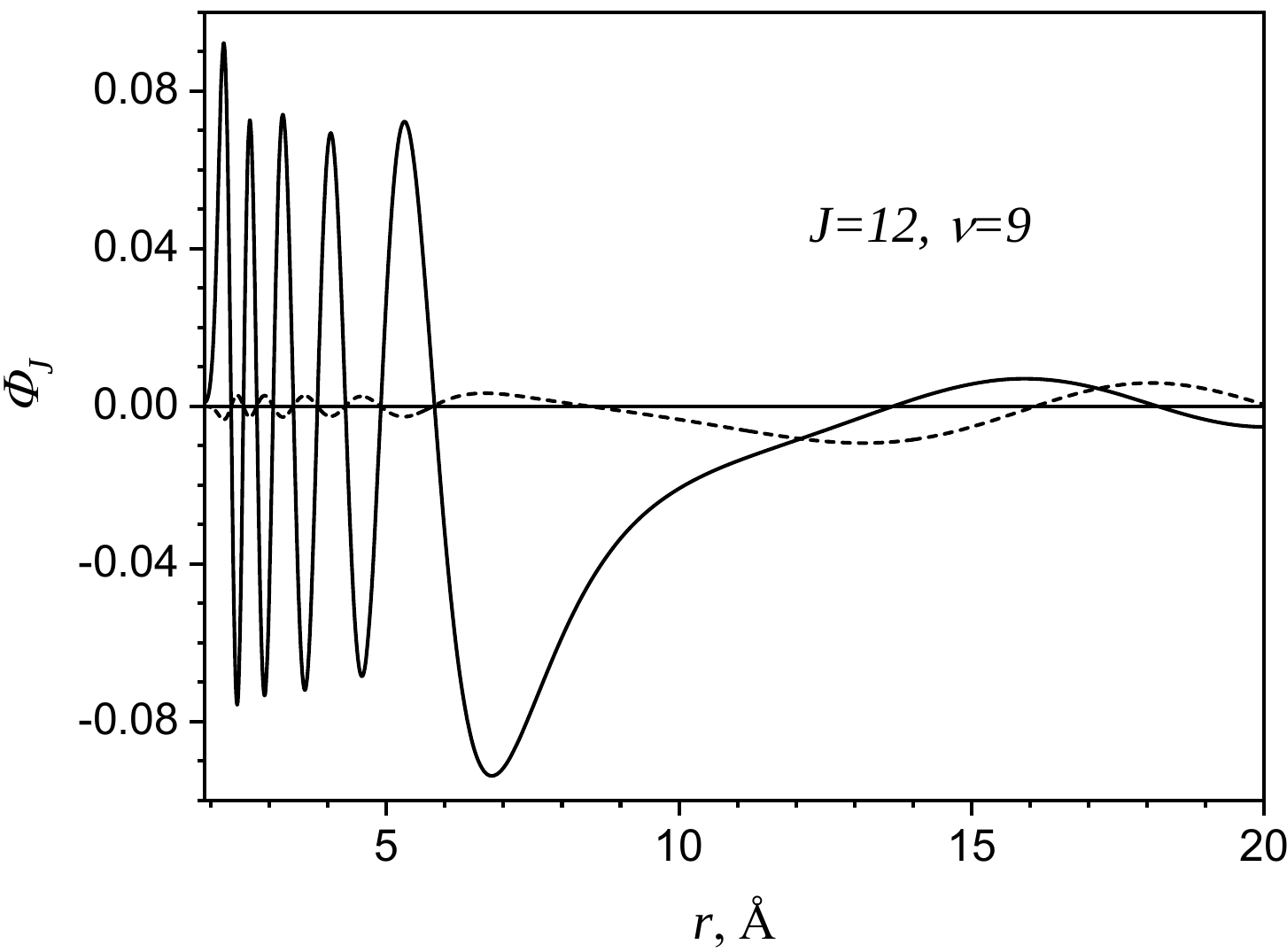}
	\includegraphics[width=0.47\textwidth]{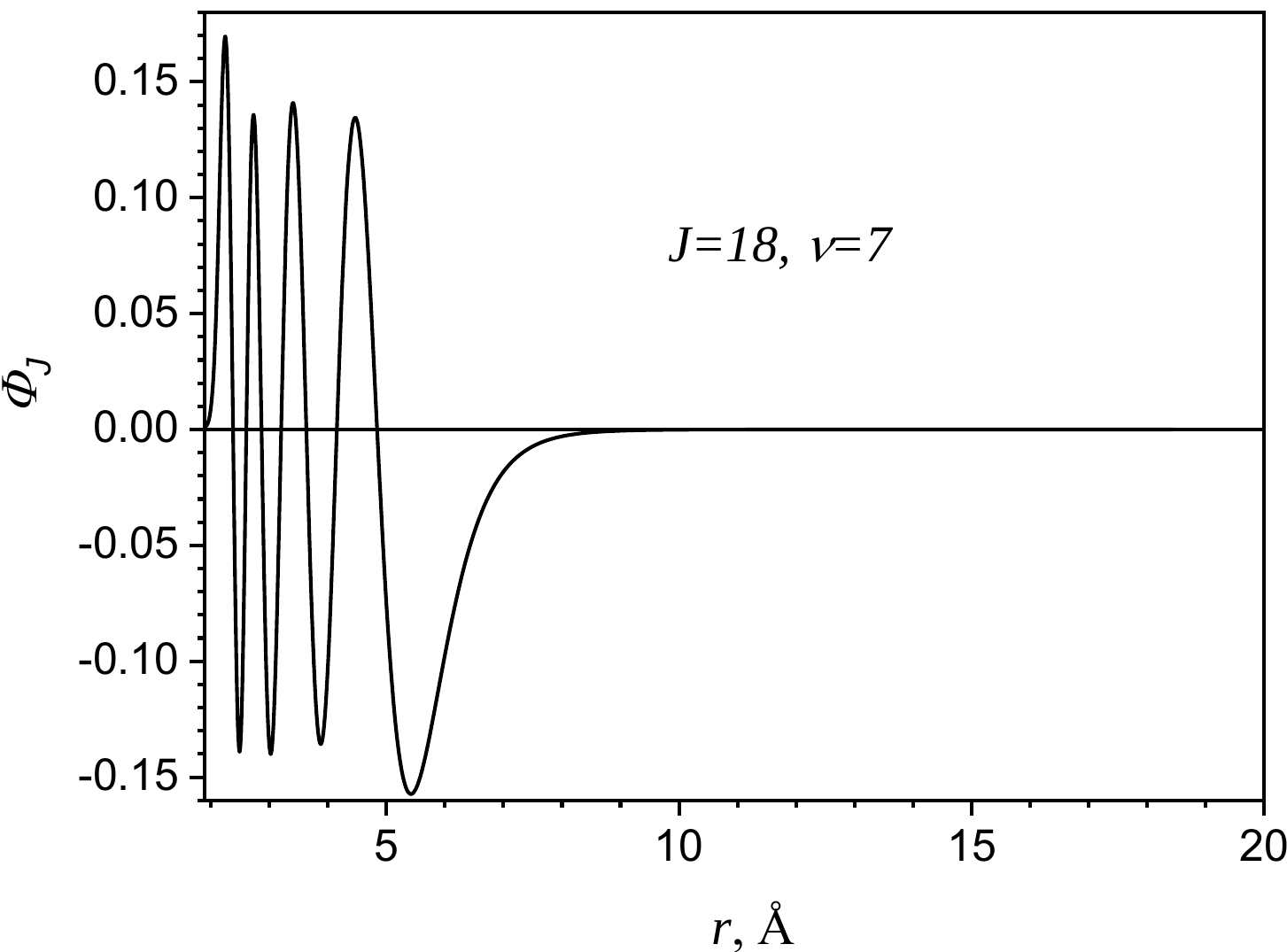}
	\includegraphics[width=0.47\textwidth]{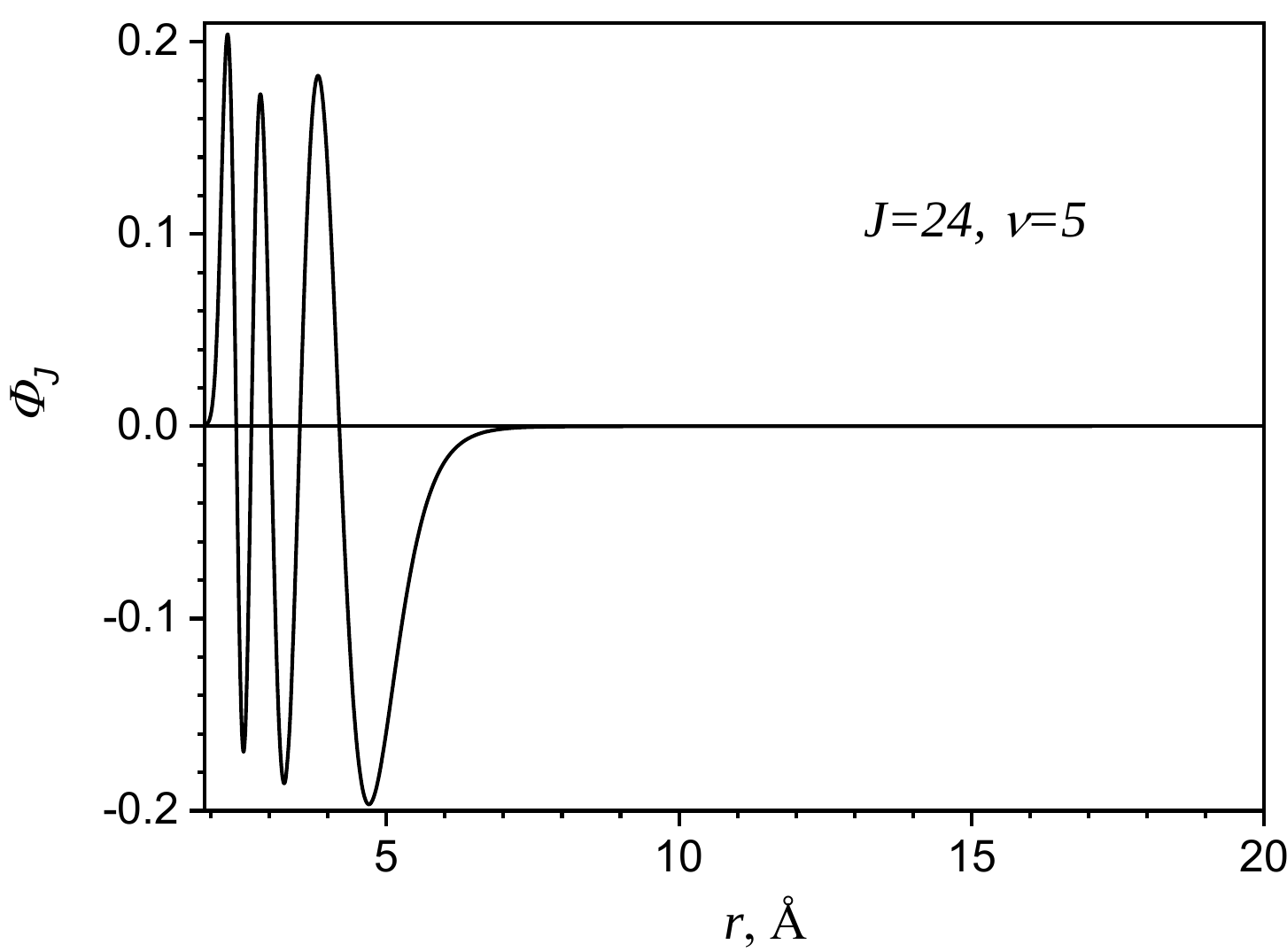}
	\includegraphics[width=0.47\textwidth]{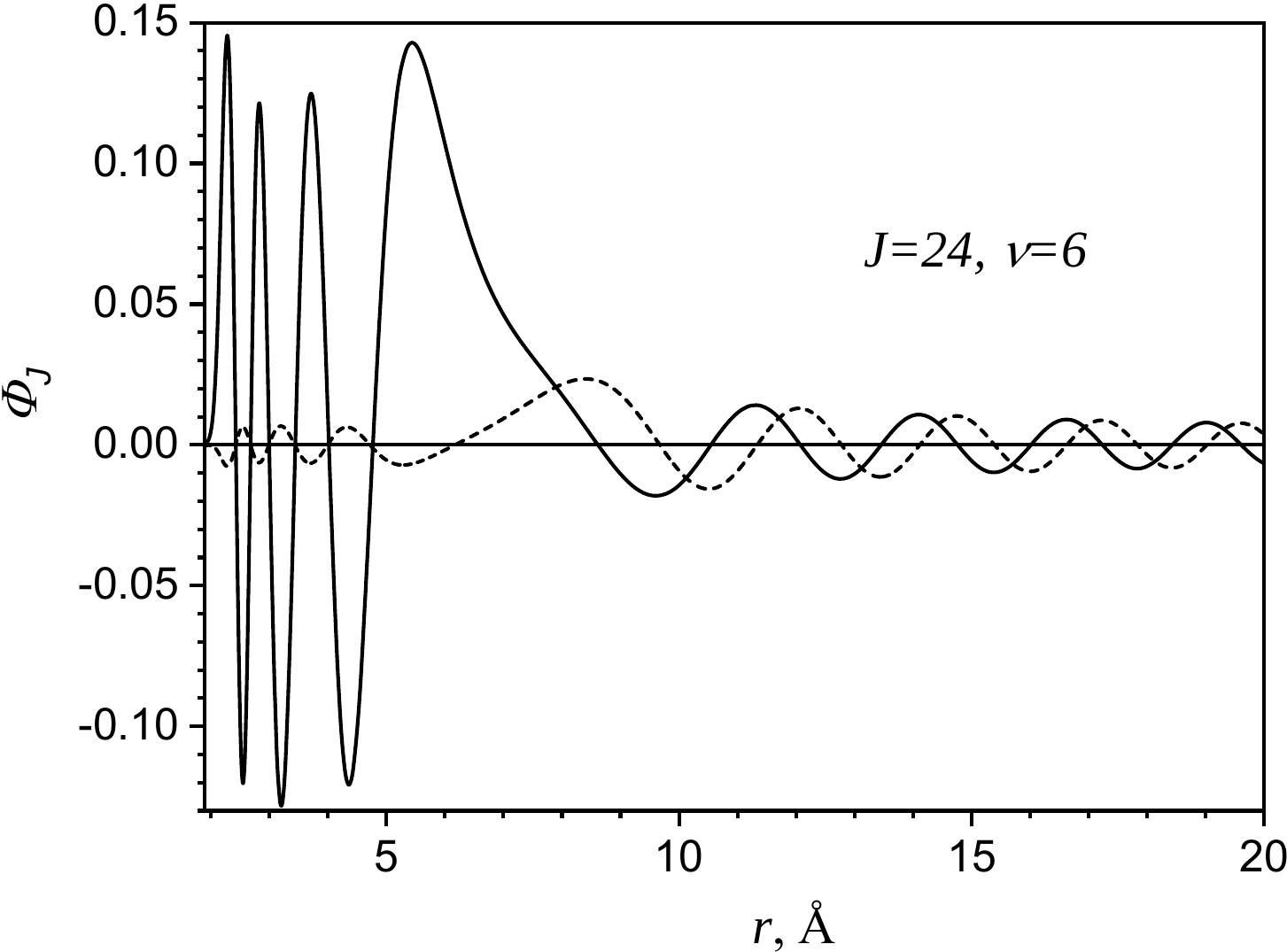}
	\caption{Plots of real (solid curve) and imaginary (dashed curve) parts of
		eigenfunctions $\Phi_{ J\nu}^M(r)$ of selected metastable states having
		eigenvalues from the Tables \ref{ms1} and \ref{ms3} marked by $J=$3, 7,
		12, 18, 24 and $\nu$.
	}\label{490}
\end{figure}
\begin{figure}[t]
	\includegraphics[width=0.47\textwidth]{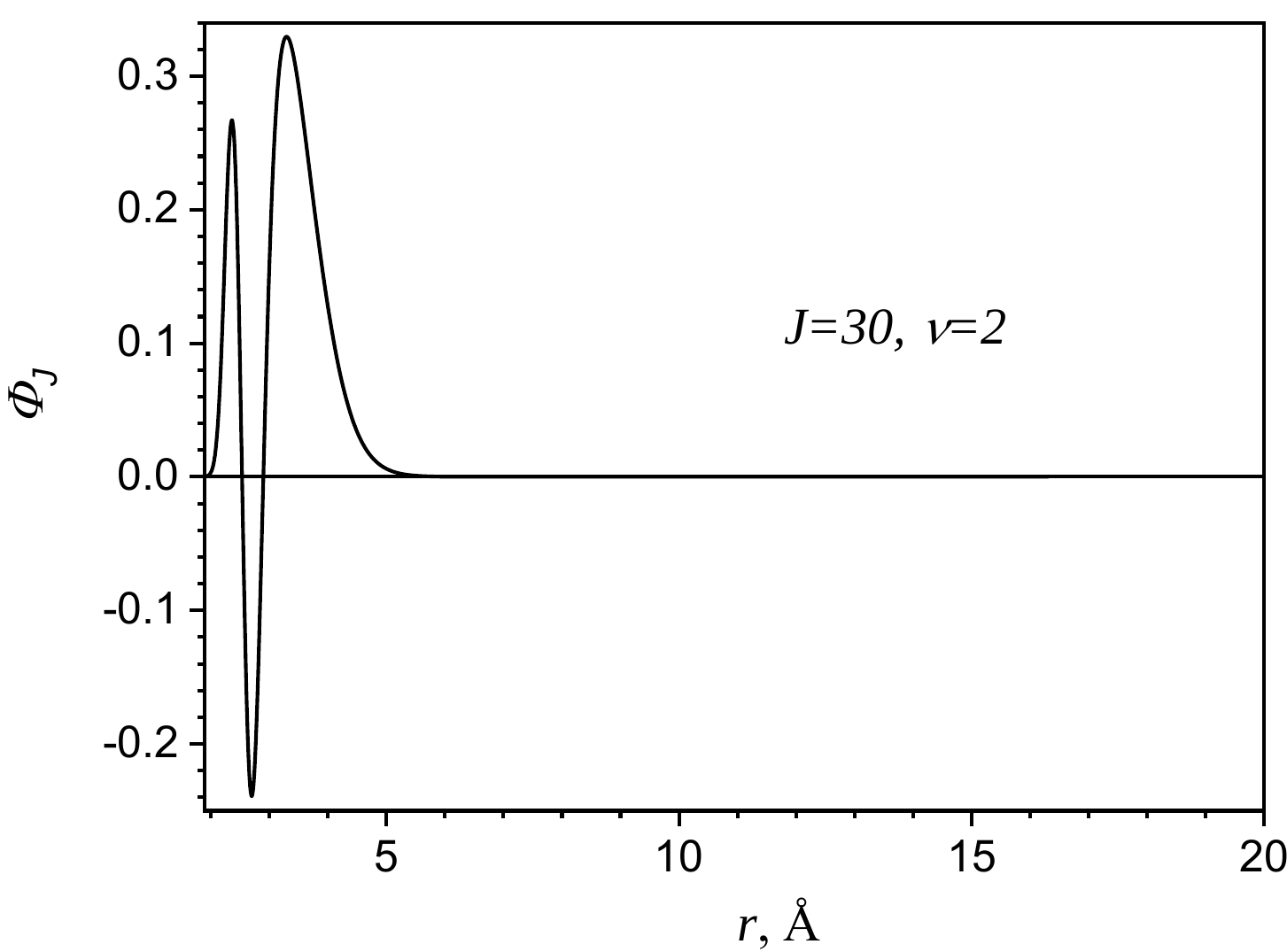}
	\includegraphics[width=0.47\textwidth]{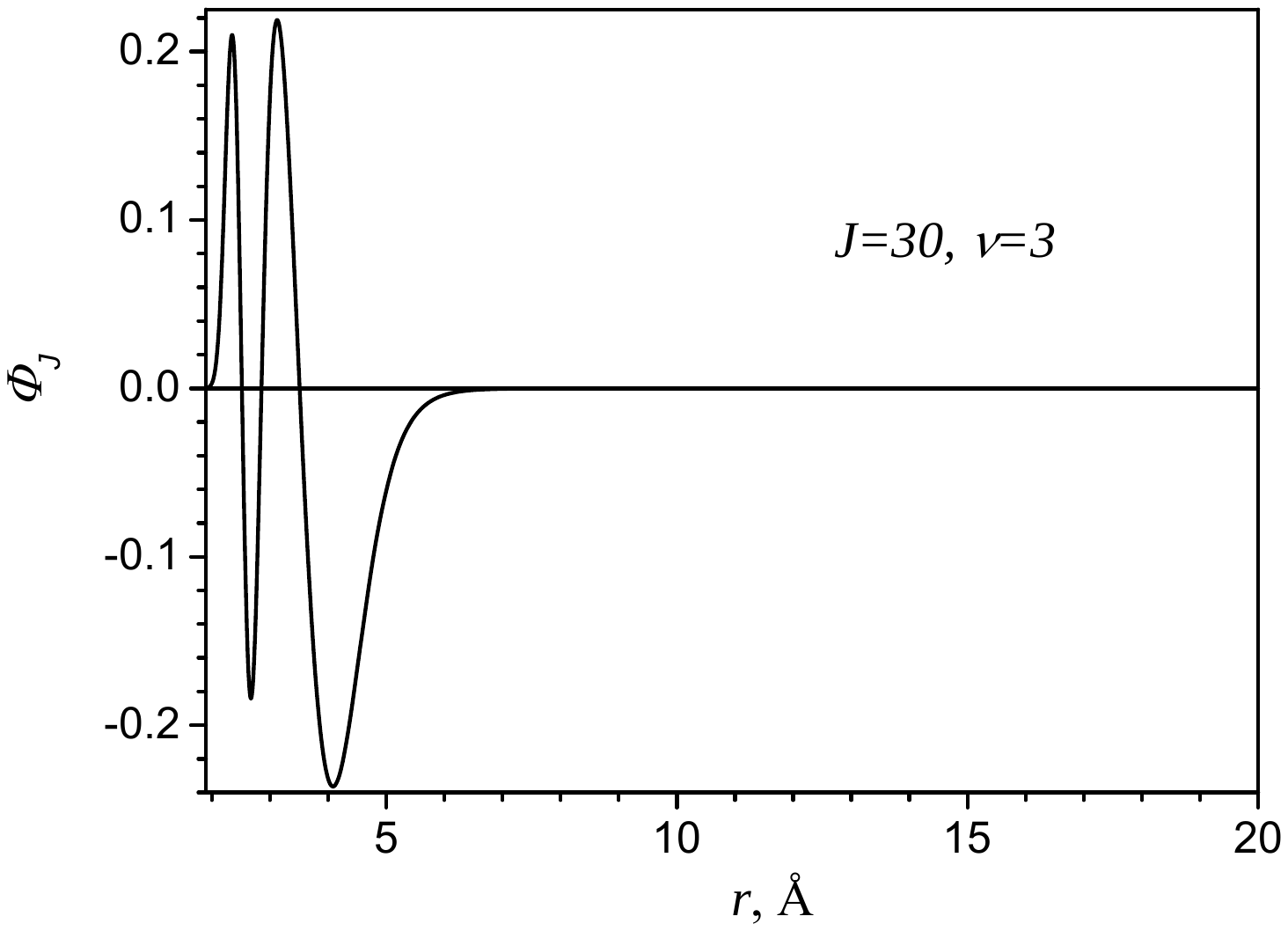}
	\includegraphics[width=0.47\textwidth]{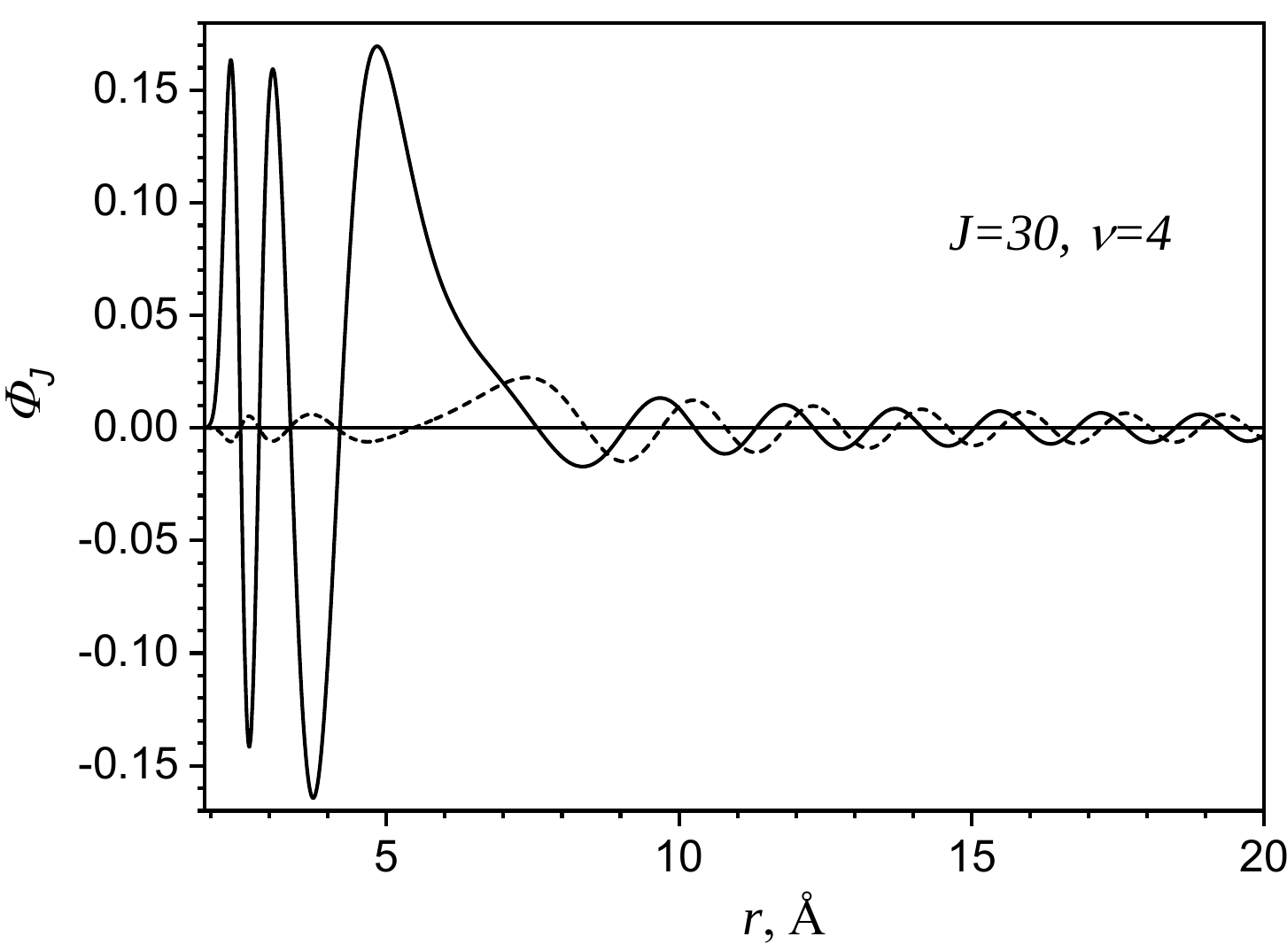}
	\includegraphics[width=0.47\textwidth]{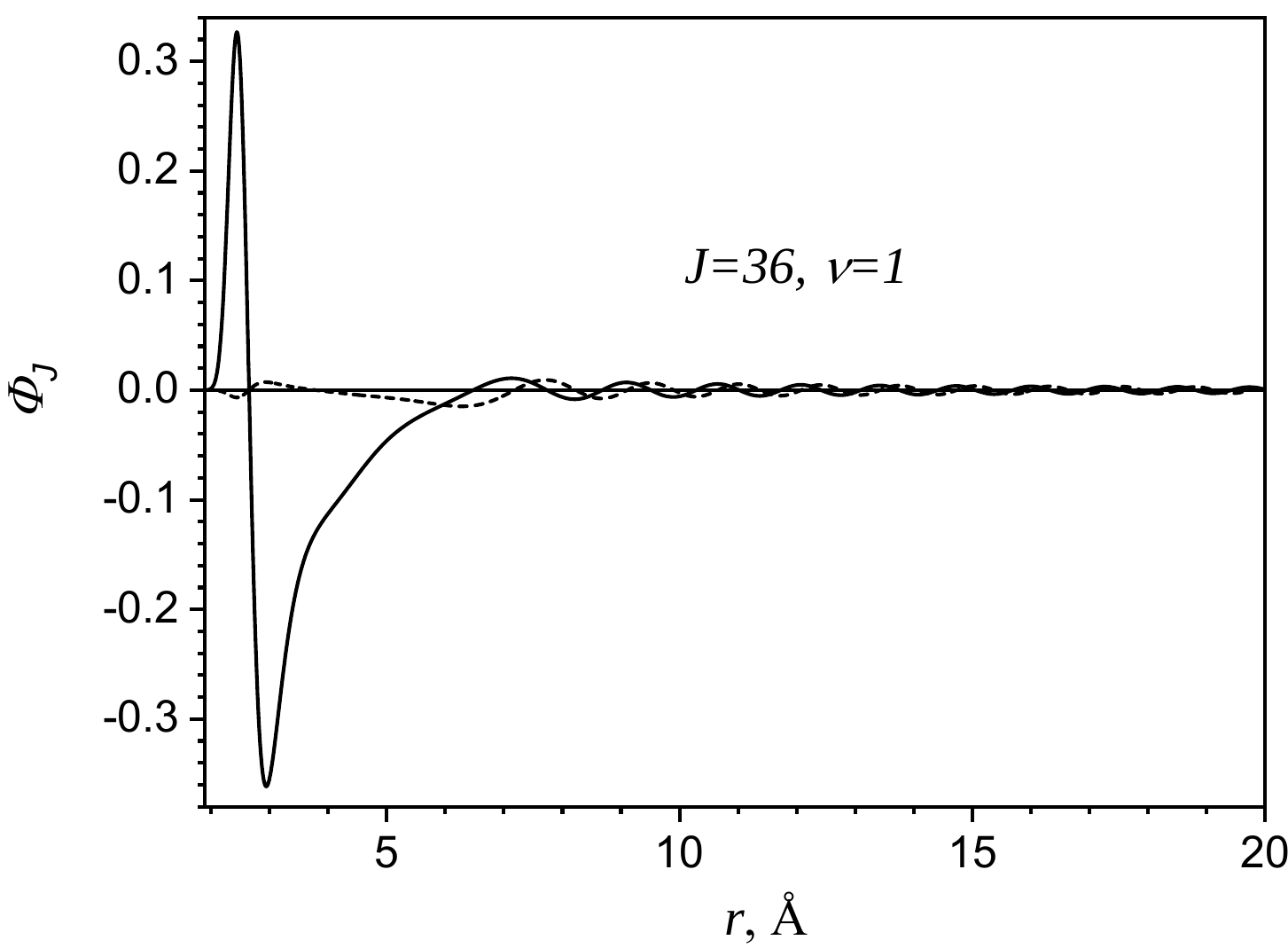}
	\includegraphics[width=0.47\textwidth]{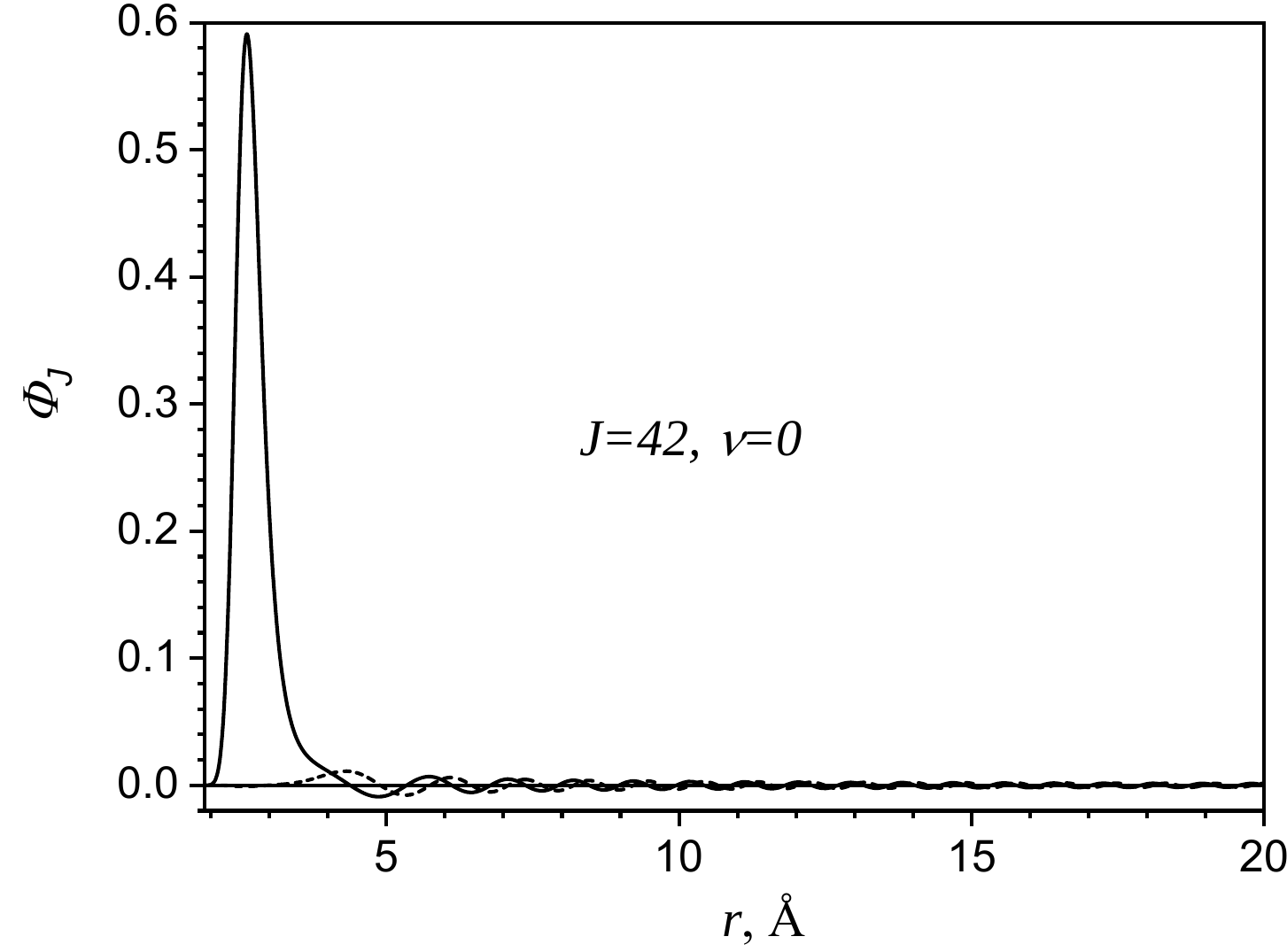}
	\includegraphics[width=0.47\textwidth]{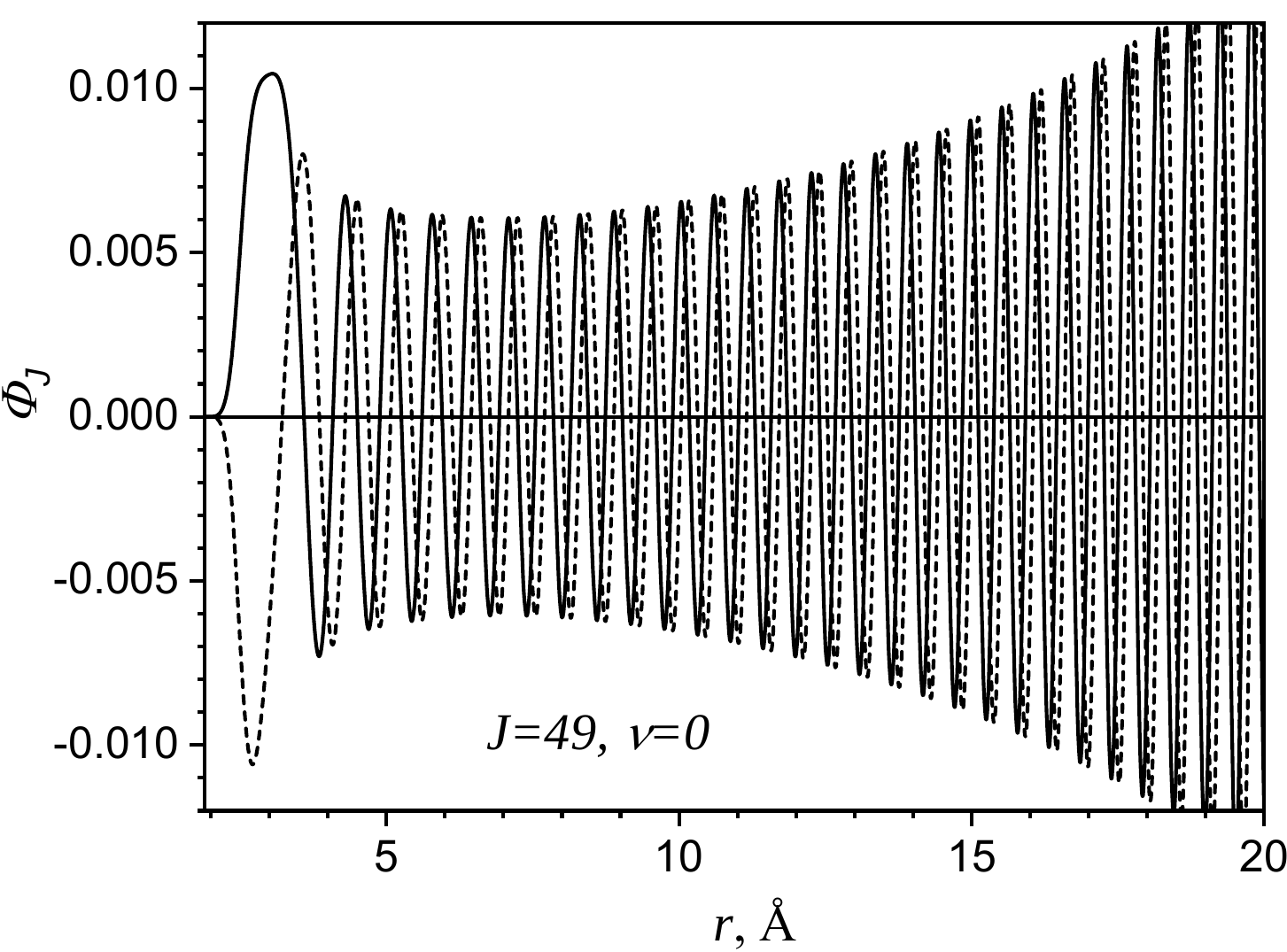}
	\caption{Plots of real (solid curve) and imaginary (dashed curve) parts of
		eigenfunctions $\Phi_{J\nu }^M(r)$ of selected metastable states having
		eigenvalues from the Tables \ref{ms3} and \ref{ms2} marked by $J=$30,
		36, 42, 49 and $\nu$.
	}\label{491}
\end{figure}

\section{Metastable states of the beryllium dimer}

In Refs. \cite{DerbovJQSRT2021_262_107529,SPIE2020}, the BVP for Eq. 
(\ref{neweqold}) was solved using the FEM program KANTBP 5M
\cite{GChuluunbaatarMC2020_1414_152} on the finite element mesh	
$\Omega_{1}{(r)}=\{1.9(0.1)2.4(0.05)2.8(0.1)20\}$ with the Neumann boundary
conditions at the boundary point $r=r_{\min}=1.90$ and the Robin boundary
condition at the boundary point $r=r_{\max}=20$ with logarithmic derivative for
$\Phi_J(kr)\equiv\Phi_{J\nu}^M(kr)$
\begin{equation}\label{logdir}
	\frac{d \Phi_J(kr)}{dr}-{\cal R}\Phi_J(kr)=0,\quad
	{\cal R}=\frac1{\Phi_{as}^+(kr)}\frac{d \Phi_{as}^+(kr)}{dr},
\end{equation}
that followed from asymptotic solution for only the outgoing wave
\cite{Kukulin1989,SiegertPR1939_56_750,GusevHaiLMCS2015_9301_182}
\begin{equation}\label{logdira}
	\Phi_{as}^+(kr)=\sqrt{k}h_{J}^{(1)}(kr)
	=-\imath\frac{\exp(+\imath (k r-\pi	J/2))}{\sqrt{k}r}+O(k^{-3/2}r^{-2}).
\end{equation}
Here complex-valued $k\equiv k^M_{J\nu}=\sqrt{{\cal E}^M_{J\nu}}=\sqrt{
E_{J\nu}^M/s_2}$ in units of \mbox{\AA}$^{-1}$\, is the wave number,
$E_{J\nu}^M=s_2{\cal E}_{J\nu}^M$ cm$^{-1}$, $h_{J}^{(1)}(z)$ is spherical
Hankel function of the first kind \cite{GoldbergerWatson}.

The complex eigenenergies $E_{J\nu}^{M}=\Re E_{J\nu}^{M}+\imath \Im
E_{J\nu}^{M}$, (in cm$^{-1}$) of Be$_2$ rota\-tional-vibrational metastable
states, where $\nu$ is the state number at a fixed value of $J$, are shown in
Tables \ref{ms1}--\ref{ms2}. Their real parts $\Re E_{J\nu}^{M}$ in comparison
with the eigenenergies $E_{\nu J}$ of vibrational-rotational bound states are
displayed in Figure \ref{l49b}a) (empty circles). This set of metastable states
is supported by the potential functions $V_J(r)$ at $J=3, 4, 7, 8 , 9, 11,
\ldots, 49$. Note that the real parts of	energies $\Re E_{J\nu}^{M}$ of the
metastable  states marked by an asterisk in	Tables \ref{ms1}--\ref{ms2} lie
above the top $V_J^{\max}$ of the potential barrier $V_J(r)$. Note that the
bound state with energy $E_{\nu=8,J=14}=-1.44$ cm$^{-1}$ for STO* PEC
corresponds to the sharp metastable state for MEMO* PEC with complex energy
$E^M_{J=14,\nu=8}=(0.083-\imath3\cdot10^{-29})$ cm$^{-1}$.

For $J>0$, the potential functions at large $r$ decrease proportionally to
$r^{-2}$ and at $J\leq38$ they have the form of a potential well with a minimum
below the dissociation threshold $D_0$, while at $J>38$ the potential well has
a minimum above the dissociation threshold. The height of the centrifugal
barrier increases with increasing $J$, but its width at the dissociation
threshold energy ($E=0$) is infinite. With increasing energy, the effective
width of the barrier decreases. The number of metastable states $\delta \nu$ at
$J\leq38$ is determined by the number of positive-energy states in the
potential well with the barrier of height  $V_J^{\max}$ taken into account,
i.e., in the well with the potential $V_J^*=\{V(r),r<r_{\max};V_{\max},r\geq
r_{\max}\}$. For small $J<16$, the barrier height $V_J^{\max}$ counted from the
zero energy is smaller than the energy difference between two upper levels of
metastable states. This means that even one metastable state can exist not for
all values of $J$. With the growth of $J$ to $J=33$, the barrier height
increases, but the width of the well changes insignificantly. As a result, the
number of metastable states increases to three. With further increase in $J$,
when in the interval $r\in (3.5,6)$ the slope of centrifugal potential exceeds
the slope of MEMO* potential, the well width rapidly decreases, so that only
two states can exist in the well, a bound state and a metastable one at $J=36$
and two metastable states at $J=34,35,37,38$. At $J\geq39$ the potential well
minimum turns to be above the dissociation threshold ($E=0$) and the effective
barrier width, the width and depth of the well decrease. Only one state exists
in the well, its width increasing with the growth of $J$. At $J>49$ there are
no energy levels in the well, and at $J>54$ the potential well disappears.

As can be seen from Tables \ref{ms1}--\ref{ms2} and Figures \ref{490},
\ref{491}, the eigenfunctions of metastable states with complex energy values
for a fixed value of the orbital momentum $J$ have an increasing number of
nodes localized inside the potential well. Beginning from each lower state
above the dissociation threshold, they have one node more than the last bound
state with real energy under the dissociation threshold $(E=0)$ with the same
value of the orbital momentum $J$ in Tables \ref{ls}--\ref{lss}. Thus, there is
a continuation of the theoretical {\it upper and lower estimates} of the real
energy eigenvalues ${ E}_{J\nu}$ to	the complex plane ${E}_{J\nu}^{M}=\Re
{E}_{J\nu}^{M}+\imath \Im {E}_{J\nu}^{M}$, labeled by the number of nodes $\nu$
of eigenfunctions localized inside the potential well, for each value of the
orbital momentum $J$.

\begin{figure}[t]
	\includegraphics[width=0.47\textwidth]{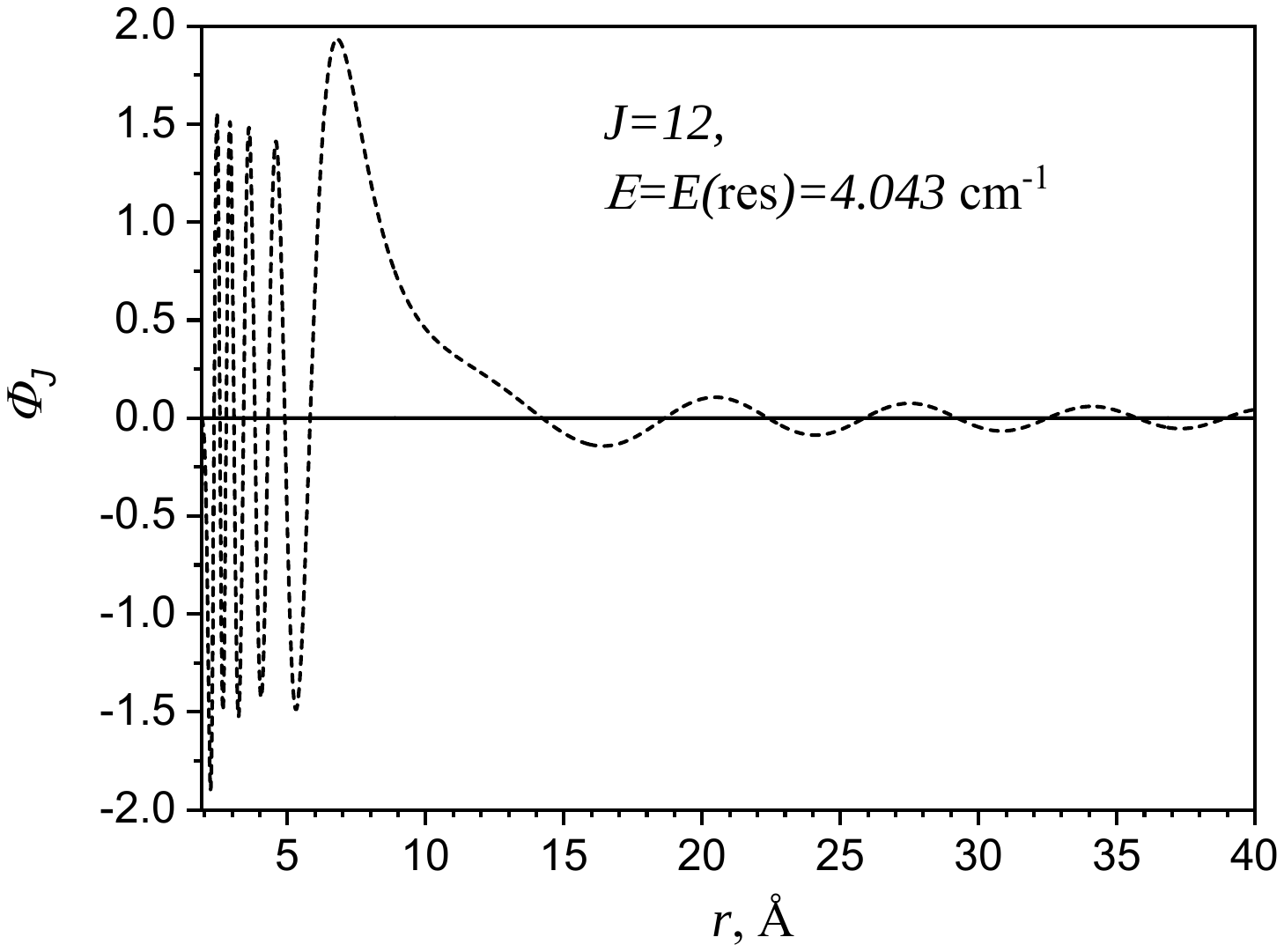}
	\includegraphics[width=0.47\textwidth]{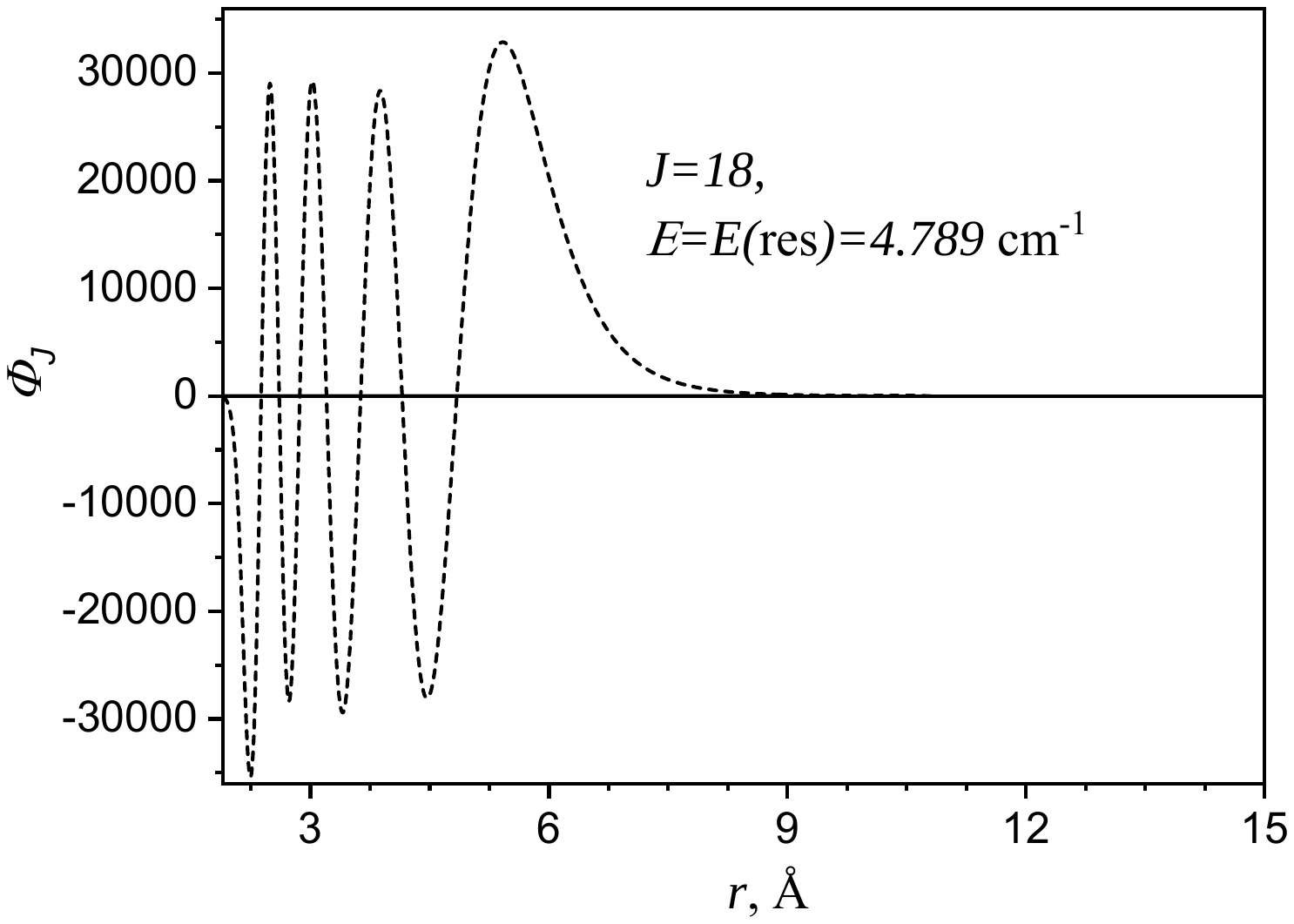}
	\includegraphics[width=0.47\textwidth]{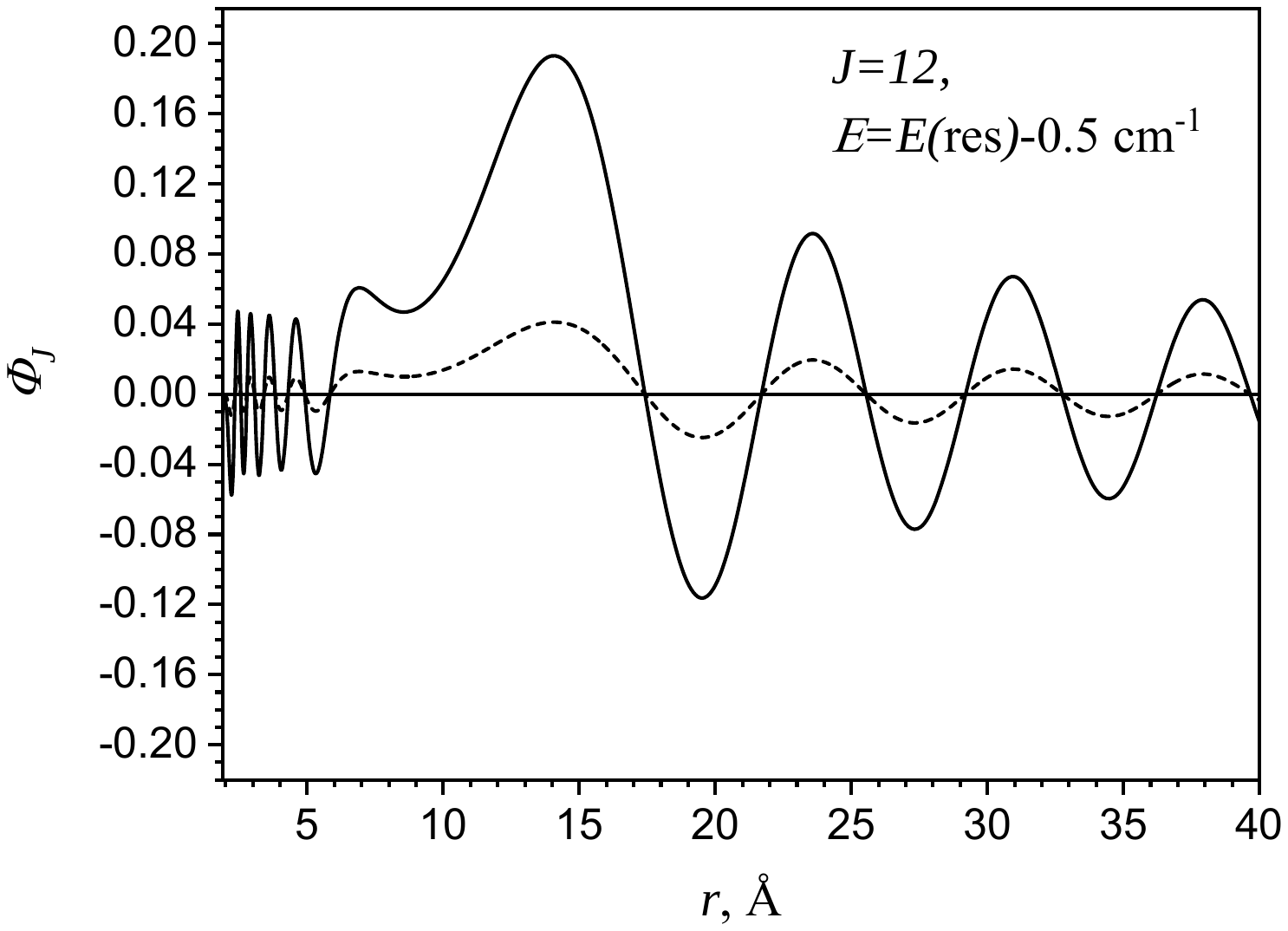}
	\includegraphics[width=0.47\textwidth]{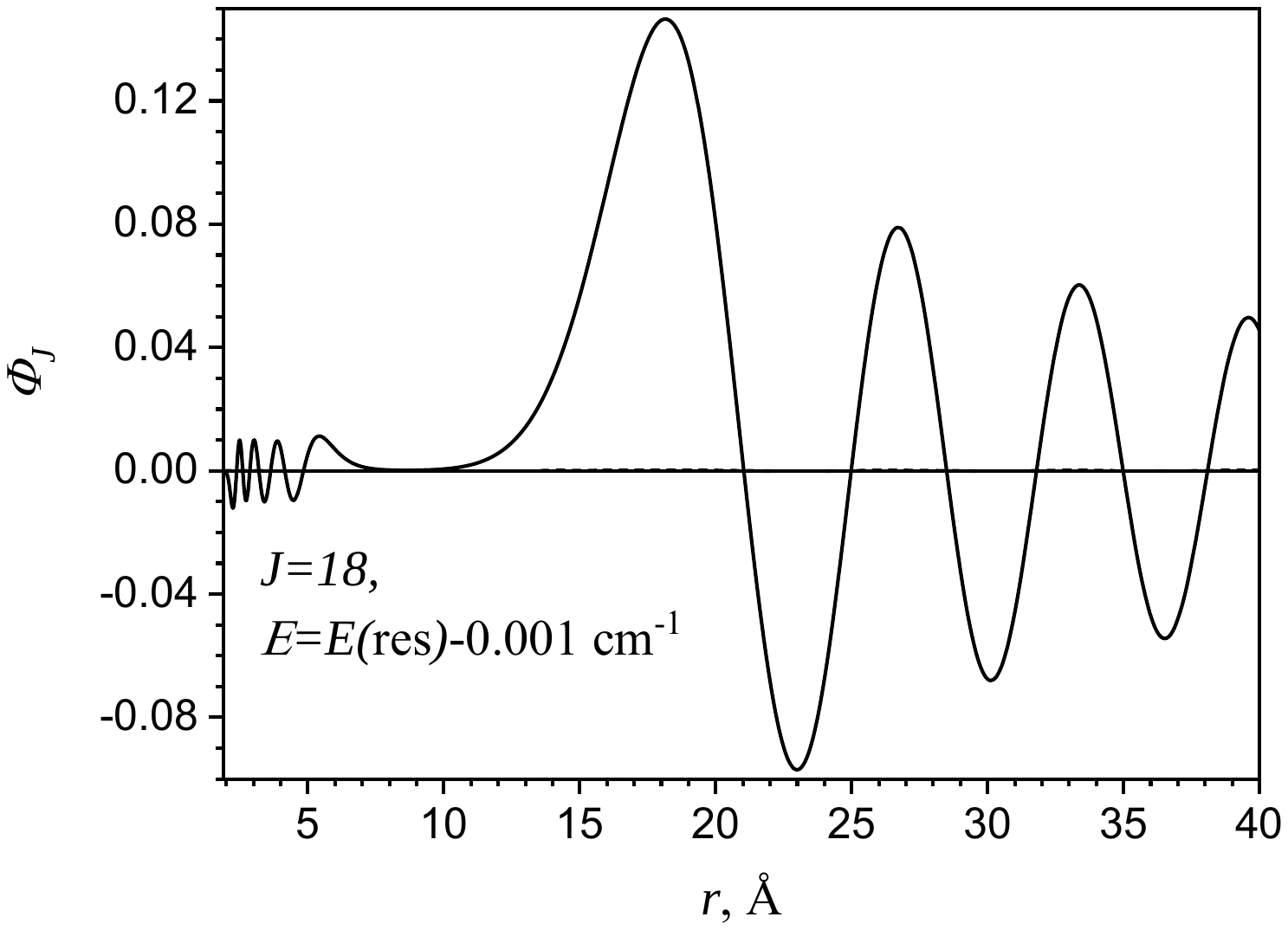}
	\includegraphics[width=0.47\textwidth]{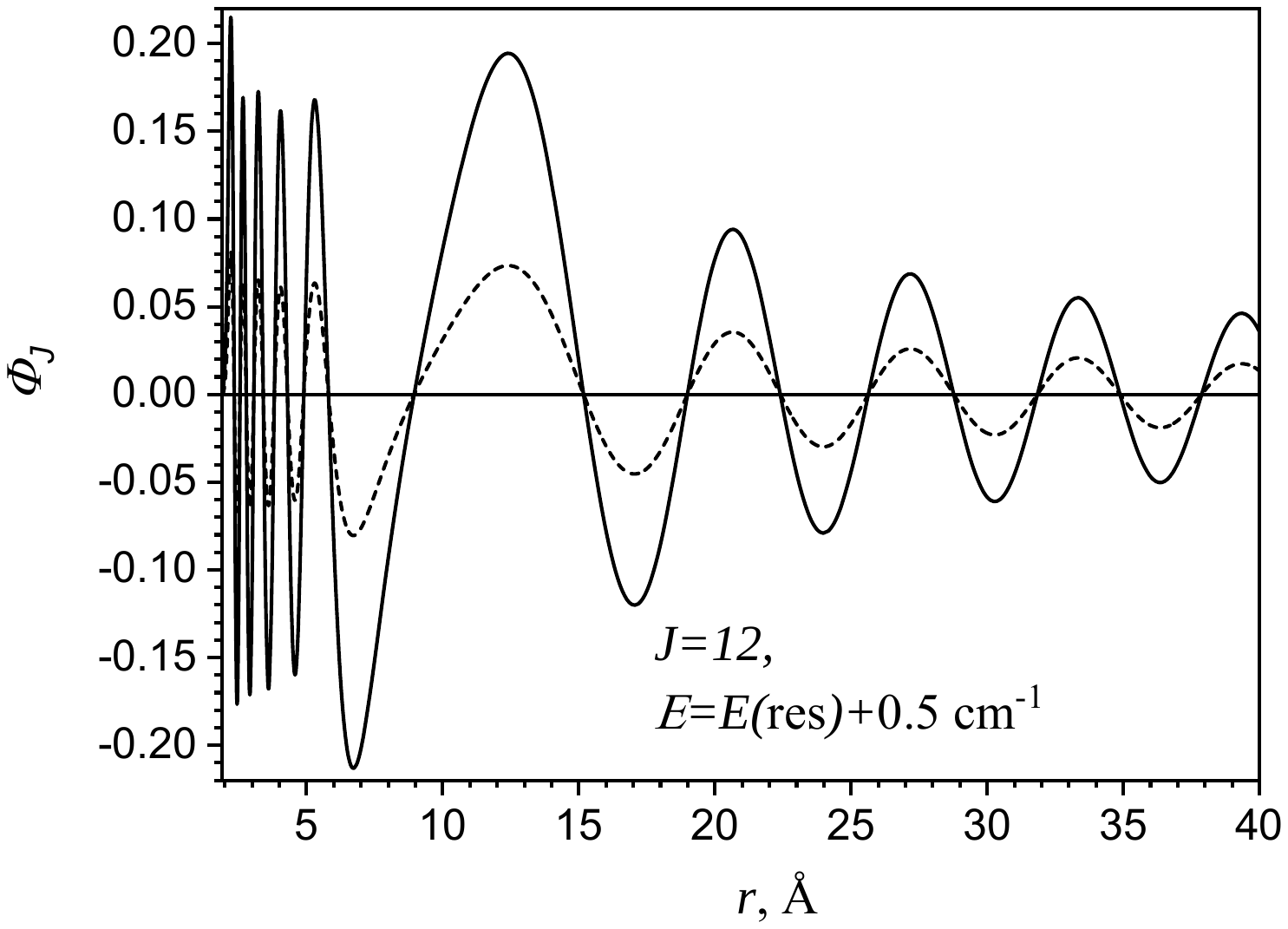}
	\includegraphics[width=0.47\textwidth]{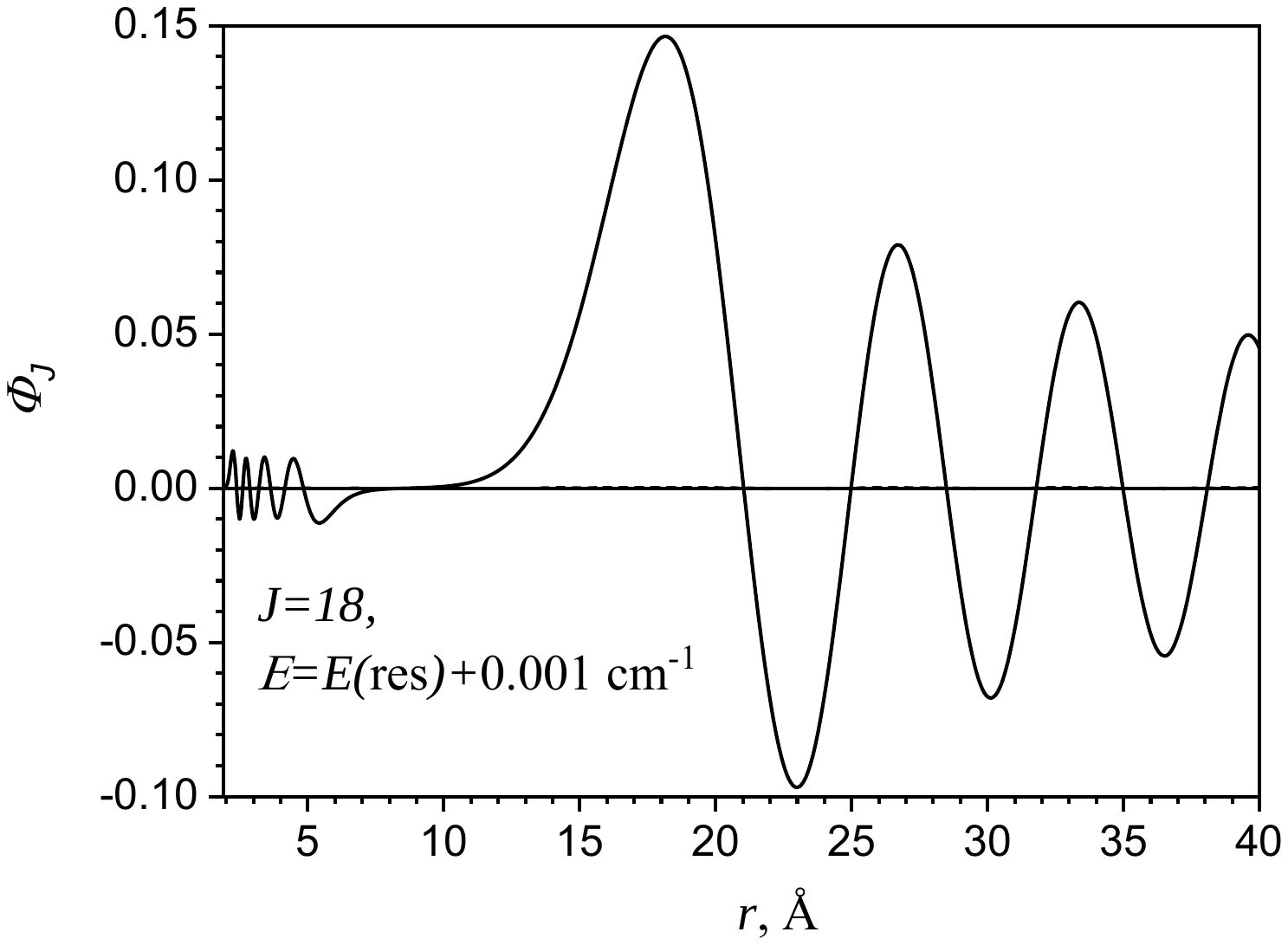}
	\caption{ Plots of the real (solid curves) and imaginary (dashed curves)
		parts of scattering wave functions $\Phi_{J}\equiv\Phi_{J}(r)$ in the
		vicinity of the resonance energy $E(res)$ at $J=12$ and $J=18$.	}
	\label{18res}
\end{figure}

\section{Scattering states of the beryllium dimer}

The scattering problem  for Eq. (\ref{neweqold}) at real-valued $E>0$ in
cm$^{-1}$ was solved using the FEM programs KANTBP 5M
\cite{GChuluunbaatarMC2020_1414_152} on the finite element mesh
$\Omega_{1}{(r)}=\{1.9(0.1)2.4(0.05)2.8(0.1)20\}$. The eigenfunctions
$\Phi_{J}(kr)$ of the scattering states are subjected to the Neumann boundary
conditions (BCs) at the boundary point $r=r_{\min}=1.90$ and the Robin boundary
condition  at the boundary point $r=r_{\max}=20$ formulated as follows:
\begin{equation}\label{logdirl}
	\frac{d \Phi_J(kr)}{dr}=\frac{d \Phi_{as}^J(kr)}{dr},\quad
	\Phi_J(kr)=\Phi_{as}^J(kr),
\end{equation}
at $r=r_{\max}$ using the asymptotic form `incident wave + outgoing wave'
\cite{GoldbergerWatson}
\begin{equation}
	\Phi_{as}^J(kr)=\frac{\imath^J}{\sqrt{2\pi}}(\Phi_{as}^-(kr)+\Phi_{as}^+(kr)S_{J}(E)).
\end{equation}
Here $S_{J}(E)=\exp(2\imath\delta_J(E))$ is the partial scattering matrix, and
the outgoing wave $\Phi_{as}^+(kr)$ and incident wave
$\Phi_{as}^-(kr)=(\Phi_{as}^+(kr))^*$ functions are given by (\ref{logdira})
at real-valued $k=\sqrt{\cal E}=\sqrt{E/s_2}>0$ in  units of \AA$^{-1}$, where
$^*$ denotes the complex-conjugation.

\begin{figure}[t]
	\includegraphics[width=0.33\textwidth,height=0.44\textwidth]{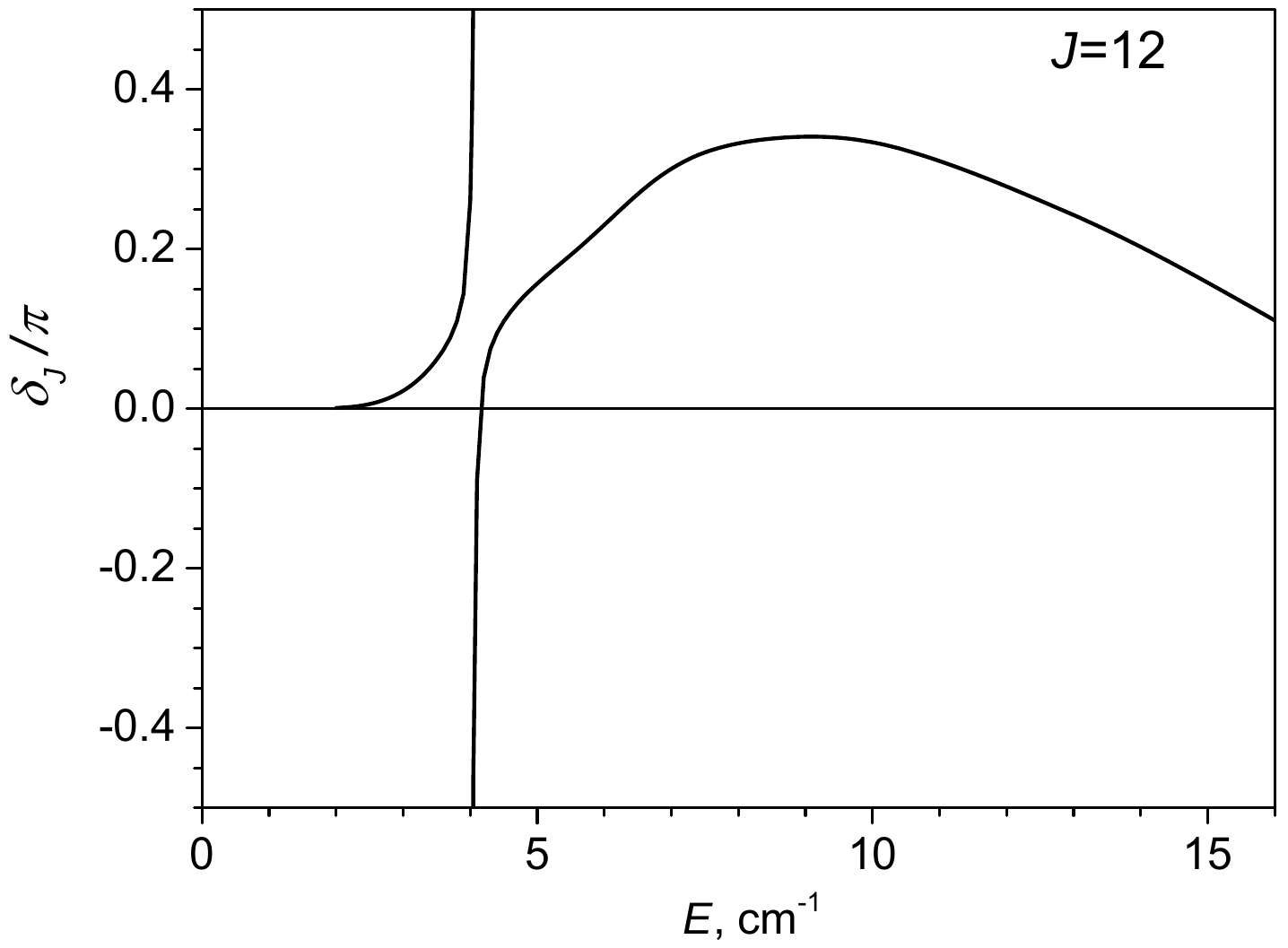}
	\includegraphics[width=0.33\textwidth,height=0.44\textwidth]{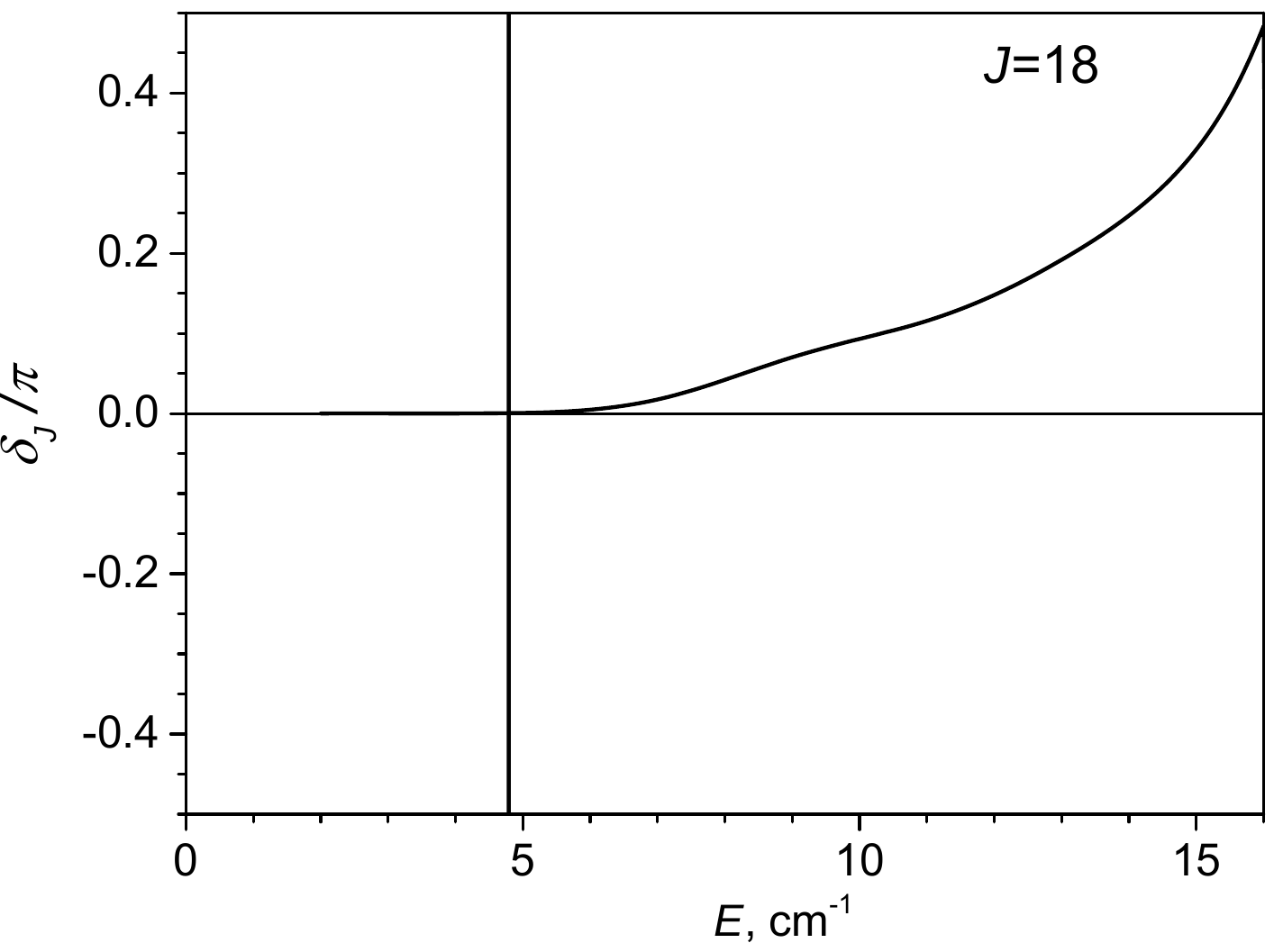}
	\includegraphics[width=0.33\textwidth,height=0.44\textwidth]{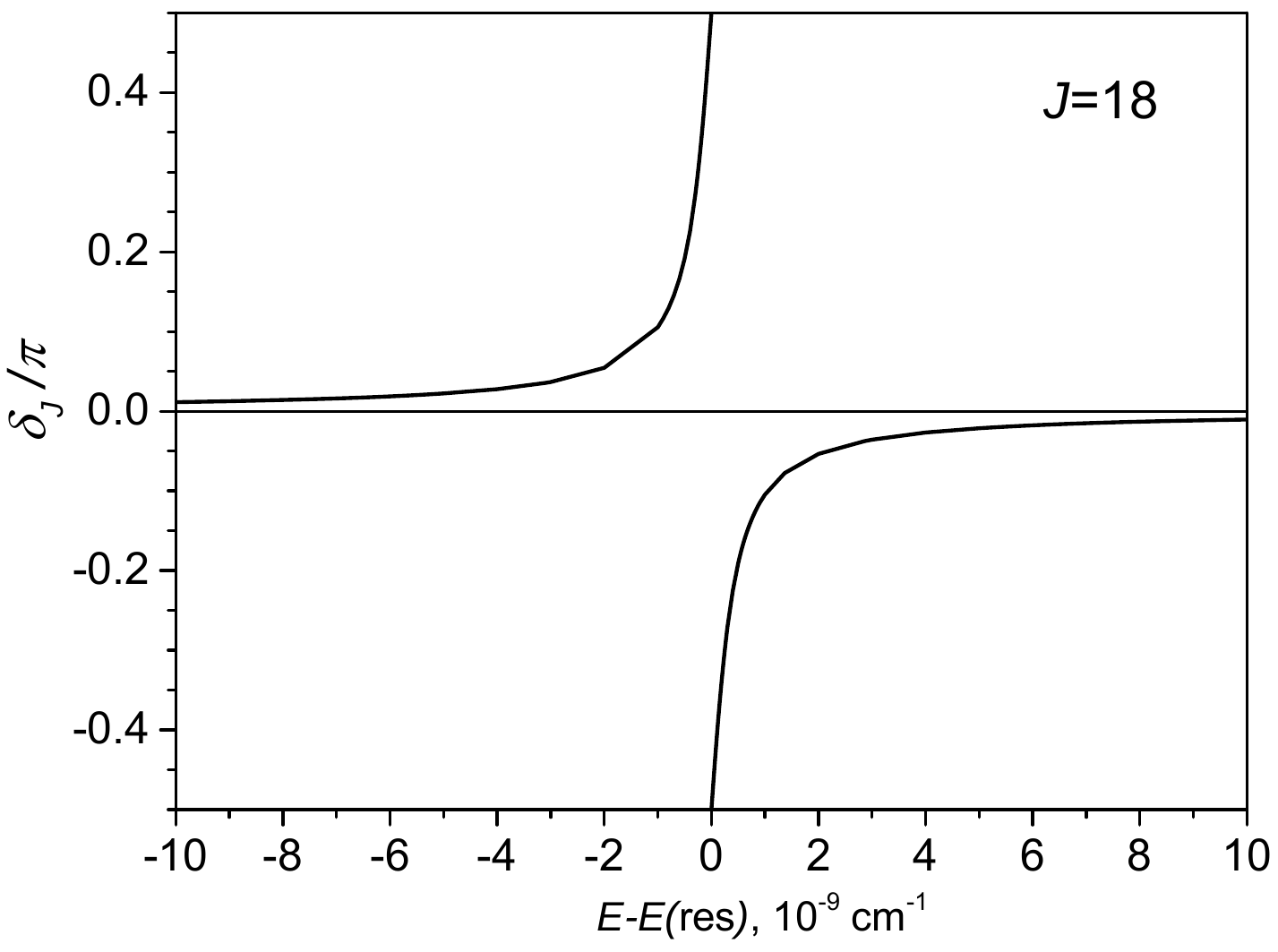}
	\caption{Phase shifts $\delta_J$ vs scattering energy $E$ counted from
	$E=0$ at $J=12$ and $J=18$.}
	\label{18sm}
\end{figure}

Plots of the real (solid curves) and imaginary (dashed curves) parts of
scattering functions in the vicinity of the resonance energy for the narrow
resonance at $J=12$ and very narrow resonance at $J=18$ are shown in Figure
\ref{18res}. One can see that the resonant scattering functions are localized
in the potential well, which is no longer observed upon a minor change in the
energy of the incident wave. As can be seen from Tables \ref{ms1}--\ref{ms2},
the energies of resonant states $E(res)$ are close to the real parts of the
energies $\Re E^{M}_{J\nu}$ of metastable states. In particular, for $J=12$
$E(res)\approx4.043$ cm$^{-1}$ and for $J=18$ $E(res)\approx4.789$ cm$^{-1}$
are close to $\Re E_{J=12\nu=9}=4.052$ cm$^{-1}$ and $\Re E_{J=18\nu=7}=4.788$
cm$^{-1}$, respectively.

In Figure \ref{18sm}, the corresponding phase shifts $\delta$ vs the scattering
energy $E$ are shown; as expected, the phase shifts take the value
$\delta=\pi/2$ for resonance energies and change rapidly in their vicinity. To
estimate the scattering length $a_0$ of the scattering state at $k\to0$ the 
lollowing  formula from Ref. \cite{GoldbergerWatson} was applied
$$a_0=-\lim_{k\rightarrow 0}
\frac{\tan\delta_0(k)}{k}\approx-\frac{d\delta_0(k)}{dk}\bigg|_{k\rightarrow0}.
$$

The recalculated value $a_0$ at $k\to0$ provides an {\it upper estimate} for
the scattering length $a_0=6.55$ \AA.

Since the MEMO* potential gives the {\it upper bound}, and the STO* potential
gives the {\it lower bound} energy of twelfth state (see Table \ref{Table_A4}),
the calculation with the STO* potential leads to the {\it lower bound} for the
scattering length $a_0=3.31$ \AA, as well as for EMO ($a_0=4.87$ \AA), MLR
($a_0=0.91$ \AA) and CPE ($a_0=0.77$ \AA) potentials
\cite{MSHHLR_JCP_2014_140_064315}.

\section{Conclusions}

The present review summarizes the results of the recent studies that show that 
the beryllium atoms are covalently bound in Be$_2$
at the low-lying vibrational energy levels with $\nu=0-4$, while at the higher
levels with $\nu=5-11$ the atoms are bonded by the van der Waals forces near
the right turning points. The EMO potential energy function used in the
experimental research \cite{MerittBondybeyHeaven_Science_2009_324_1548} for
fitting the measured vibrational energy levels of Be$_2$ does not correctly
describe this dual nature of the chemical bonding. It describes better the
covalent bonding on the low-lying vibrational energy levels than the van der
Waals bonding on the upper ones. A comparison of the EMO potential energy
function with the potential function obtained in the high precision \textit{ab
initio} calculation carried out in the present study shows that the EMO
function is too narrow near the dissociation limit. Therefore, the modified EMO
potential function has been constructed by replacing the parts of the original
EMO function near dissociation limit and above it by the \textit{ab initio}
potential function. The obtained MEMO potential function not only has the
correct dissociation energy, but, in distinction with EMO potential function,
also describes all twelve vibrational energy levels with a smaller RMS error of
less than 0.4 cm$^{-1}$.

Special attention the papers reviewed was focused on improved calculations of 
spectrum of the bound vibrational-rotational state together with spectrum of the
metastable vibrational-rotational state having complex-valued eigenenergies.
The existence of these metastable states is confirmed by calculation of the
corresponding scattering states with real values of the resonance energies.
Theoretical {\it upper and lower estimates} are of significant importance for
further experiments in laser spectroscopy of the beryllium dimer. It is also
important for modeling of a near-surface diffusion of the beryllium dimers
\cite{PenkovKrass2014_47_225210, Scripta14, Vinitsky2014_8660_472,
GusevTMF2016_186_21, polonika17, GusevPhAN2018_81_945} in connection with the
well-known multifunctional use of beryllium alloys in modern technologies of the
electronic, space and nuclear industries \cite{WalshKA2009}, and, in
particular, the ITER project \cite{GAlloucheJP2008_117_012002}.

\section{Acknowledgements}

This publication has been supported by the Russian Foundation for Basic
Research and Ministry of Education, Culture, Science and Sports of Mongolia
(the grant 20-51-44001) and the Peoples' Friendship University of Russia (RUDN)
Strategic Academic Leadership Program, project No.021934-0-000. This research
is funded by Ho Chi Minh City University of Education Foundation for Science
and Technology (grant No. CS.2021.19.47). OCH acknowledges financial support
from the Ministry of Education and Science of Mongolia (grant No. ShuG
2021/137).

\section{References}

\bibliography{manuscript_ref}

\end{document}